\documentclass[12pt,letterpaper]{article}     
\usepackage[skins,theorems]{tcolorbox}
\tcbset{highlight math style={enhanced,
  colframe=red,colback=white,arc=0pt,boxrule=1pt}}
  \usepackage[bookmarksopen, bookmarksnumbered, bookmarksopenlevel=2]{hyperref}
  \usepackage{tikz}
  \usepackage{tikz-3dplot}
 \usetikzlibrary{calc}
 \usetikzlibrary{decorations} %
 \usepackage[toc,page]{appendix}
 \usepackage{amsmath}
 \usepackage{amssymb}
 \usepackage{graphicx}
 \usepackage{hhline}
 \usepackage[bf]{caption}
\usepackage{cite}
\usepackage{booktabs}
\usepackage{empheq}
\usepackage{paralist}
\usepackage[vcentermath]{youngtab}
\usepackage{geometry}
\usepackage{slashed}
\usepackage{color}
\usepackage{stackrel}
\usepackage{tikz-cd} 
\usepackage{mathtools}
\usepackage{cancel} 
\usepackage{multirow}
\usepackage{physics}

\usepackage{empheq}
\usepackage{arydshln}

\newcommand{\bqa}{\begin{eqnarray}}
\newcommand{\eqa}{\end{eqnarray}}

\newenvironment{eqn}{\begin{equation}\begin{aligned}}{\end{aligned}\end{equation}\noindent}
\newenvironment{eqn*}{\begin{equation*}\begin{aligned}}{\end{aligned}\end{equation*}\noindent}
\hypersetup{
    pdftitle={},
    pdfauthor={},
    pdfsubject={}
}
\numberwithin{equation}{section}
\numberwithin{table}{section}

\makeatletter

\newcommand{\be}{\begin{equation}}
\newcommand{\ee}{\end{equation}}
\newcommand{\beq}{\begin{equation}}
\newcommand{\eeq}{\end{equation}}
\newcommand{\ba}{\begin{aligned}}
\newcommand{\ea}{\end{aligned}}

\newcommand{\bea}{\begin{eqnarray}}
\newcommand{\eea}{\end{eqnarray}}

\newcommand{\cA}{\mathcal{A}}

\newcommand{\cF}{\mathcal{F}}

\newcommand\bi{\begin{itemize}}
\newcommand\ei{\end{itemize}}

\def\Im{\mathop{\mathrm{Im}}\nolimits}

\def\Re{\mathop{\mathrm{Re}}\nolimits}

\def\Tr{\mathop{\mathrm{Tr}}\nolimits}

\def\unit{{1\kern-.65ex {\rm l}}}
\def\1{{1\kern-.65ex {\rm l}}}

\def\bbC{{\mathbb{C}}}

\def\bbN{{\mathbb{N}}}

\def\bbR{{\mathbb{R}}}

\newcommand{\Rg}{\mathbb{R}_{\mathscr{G}}}
\newcommand{\Rge}{\mathbb{R}_{\mathscr{G}, \exp}}

\newcount\hour \newcount\minute
\hour=\time \divide \hour by 60
\minute=\time
\def\now{%
\ifnum \hour<13
  \ifnum \hour=0 \advance \hour by 12 \number\hour:\else \number\hour:\fi%
     \ifnum \minute<10 0\fi%
     \number\minute%
\ A.M.%
\else \advance \hour by -12 \number\hour:%
  \ifnum \minute<10 0\fi%
  \number\minute%
  \ P.M.%
\fi%
}

\makeatother

 \usepackage{mathrsfs} 
\usepackage{caption}
\usepackage{subcaption}

\begin{document}

\begin{titlepage}
\begin{center}
\rightline{\small }

\vskip 15 mm

{\large \bf

} 
\vskip 11 mm

\begin{center}
{\Large \bfseries Taming non-analyticities of QFT observables}~\\[.3cm]

\vspace{1cm}
{\bf Thomas W. Grimm}, {\bf Giovanni  Ravazzini}, {\bf Mick van Vliet}

\vskip 11 mm 
{\small
Institute for Theoretical Physics, Utrecht University\\ Princetonplein 5, 3584 CC Utrecht, The Netherlands\\[3mm]
}

\vspace*{1.5em}

\end{center}

\vskip 11 mm

\end{center}
\vskip 17mm

\begin{abstract}
\noindent
Many observables in quantum field theories are involved non-analytic functions of the parameters of the theory. However, it is expected that they are not arbitrarily wild, but rather have only a finite amount of geometric complexity. This expectation has been recently formalized by a tameness principle: physical observables should be definable in o-minimal structures and their sharp refinements. In this work, we show that a broad class of non-analytic partition and correlation functions are tame functions in the o-minimal structure known as $\mathbb{R}_{\mathscr{G}}$ -- the structure defining Gevrey functions. 
Using a perturbative approach, we expand the observables in asymptotic series in powers of a small coupling constant. 
Although these series are often divergent, they can be Borel-resummed in the absence of Stokes phenomena to yield the full partition and correlation functions.
We show that this makes them definable in $\mathbb{R}_{\mathscr{G}}$ and provide a number of motivating examples. 
These include certain 0-dimensional quantum field theories and a set of higher-dimensional quantum field theories that can be analyzed using constructive field theory.
Finally, we discuss how the eigenvalues of certain Hamiltonians in quantum mechanics are also definable in $\mathbb{R}_{\mathscr{G}}$.

\end{abstract}

\end{titlepage}

\newpage

\tableofcontents
\newpage

\section{Introduction}

The observables of quantum field theories (QFTs) are typically defined as path integrals over field configurations and generally depend in a non-trivial way on the parameters of the Lagrangian of the theory. While these functions are often extremely hard to compute, it appears reasonable to presume that they are not arbitrarily complex and have many regularity features. Specifically, one can expect that they are specified by inputting a finite amount of information that might be matched with experimental data. While such a general principle is appealing, the major challenge is to formulate it as a well-defined mathematical statement. One might hope, for example, to single out classes of functions that are sufficiently `tame' and admit a well-defined measure of complexity or information. Recently, a set of concrete proposals was introduced in \cite{Grimm:2021vpn,Douglas:2022ynw,Douglas:2023fcg,Grimm:2023xqy}, suggesting that the appropriate tameness property for observables is their definability within o-minimal structures \cite{MR1633348} and their sharp refinements \cite{beyondomin,binyamini2022sharplyominimalstructuressharp,newBinyamini}. The main goal of this work is to make concrete steps to test and solidify these proposals for certain well-studied physical QFT observables.

To make tameness statements about observables, we need to restrict ourselves to situations in which they can be reliably analyzed. 
The most common approach to calculating a QFT observable $\cA(\lambda)$ is perturbation theory, in which a small coupling parameter $\lambda$ is used to perform a series expansion. The individual terms of this perturbative expansion may then be computed by means of Feynman diagram methods. 
However, as readily seen by the factorial growth of the number of Feynman diagrams, the resulting series expansion often diverges and is only asymptotic to the observable $\cA(\lambda)$. This prevalent feature of QFTs has been 
recognized since the seminal work of Dyson \cite{PhysRev.85.631}. 
Another manifestation of this behavior is the non-analyticity of observables as $\lambda$ approaches zero, which poses a significant challenge to using perturbation theory to test the tameness of the full quantum observable. Nevertheless, we expect this feature to prevail. Even though $\cA(\lambda)$ will gain contributions from infinitely many virtual processes in the perturbative expansion, the final result should still be tame and describable with only a finite amount of information.

A powerful method of addressing the divergence of the perturbative expansions in QFT is Borel resummation, which is a procedure that transforms the asymptotic series arising from perturbation theory into a function of $\lambda$. In this work we will focus on perturbative expansions that have no Stokes phenomena. More precisely, we constrain ourselves to series that are Borel-summable, or $p$-summable, which means that the Borel transform of the series has no poles on the positive real line. Borel summability implies that we will not need to promote the perturbative series to a transseries including non-perturbative instanton sectors. This simplifying assumption guarantees that there is no additional non-perturbative information that we need to supply to recover the function $\cA(\lambda)$. 
Under this assumption, we build on a powerful theorem that implies that the considered partition and correlation functions, although not analytic in the weak coupling-limit, are quasi-analytic functions. This fact will be a key ingredient to show the tameness of observables. Concretely, we will prove that partition and correlation functions in certain QFTs are definable functions in the o-minimal structure generated by the Gevrey functions, known as $\Rg$ \cite{DRIES_SPEISSEGGER_2000}. Gevrey functions are smooth functions whose non-analytic  behavior at the origin exactly matches that of the full observable.

It is interesting to compare our findings with the claims and observations made in earlier works \cite{Grimm:2021vpn,Douglas:2022ynw,Douglas:2023fcg,Grimm:2023xqy}. In reference \cite{Douglas:2022ynw} it was proved that the perturbative expansion of scattering amplitudes, truncated at any finite order, are definable in the o-minimal structure $\bbR_{\textup{an}, \exp}$ as functions of masses, external momenta and the couplings. This statement is immediate when viewing the amplitudes only as function of the couplings, since at finite loop-order they appear polynomially. In this work, we find that $\bbR_{\textup{an}, \exp}$ no longer suffices when considering the full perturbative expansion in the couplings. While $\Rg$ contains $\bbR_{\rm an}$, it allows for defining in addition the non-analytic functions needed to recover the full perturbative expansion. We thus make a concrete next step in the search for the o-minimal structure relevant for physical systems, as set out in \cite{Douglas:2023fcg}. Whether or not there will eventually be a unique structure for physics remains open, since it is already known that no largest o-minimal structure exists \cite{DenjoyCarleman}. Furthermore, if one insists that the o-minimal structure possesses a notion of complexity or information, as proposed in \cite{Grimm:2023xqy}, then $\bbR_{\rm an},\bbR_{\rm an,exp}$ and $\Rg, \bbR_{\mathscr{G},\text{exp}}$ are not yet suitable \cite{binyamini2022sharplyominimalstructuressharp}. It is an exciting recent direction in mathematics research to find suitable replacements admitting such a notion \cite{newBinyamini}.

The outline of this work is as follows. After a brief introduction to non-analyticity and Borel summability in section \ref{sec:nonanal-Borel}, we proceed by defining o-minimal structures and tame functions in section \ref{o-minimality-section}. A special focus will lie on introducing the structure $\Rg$. 
In section \ref{0-dim} we show that some partition functions and correlation functions for QFTs in zero dimensions are tame functions in $\Rg$. Thereafter, in section \ref{high-dim}
we argue for the tameness of partition functions in higher-dimensional QFTs by using results from constructive field theory. Finally, we comment briefly on how asymptotic series can also be used to  establish the tameness of the energy eigenvalues of a perturbed Hamiltonian. Similar to amplitude power series, these must be Borel-resummed: we discuss examples in which this is possible, thereby resulting in a tame function of $\Rg$, and counterexamples in which Borel resummation is impaired by the Stokes phenomenon.

\section{Non-analytic partition functions and Borel sums} \label{sec:nonanal-Borel}

In this section we illustrate how QFT partition functions $Z(\lambda)$ are generically non-analytic functions in the weak coupling limit $\lambda \rightarrow 0$. The non-analyticity manifests itself in the divergence of the perturbative expansion in $\lambda$. To nevertheless use this expansion to capture information about $Z(\lambda)$, we are then led to introduce the Borel resummation procedure. This technique allows one to turn the divergent series %can be 'summed' 
into a well-defined function that agrees with $Z(\lambda)$ in the absence of resurgence phenomena. This section will set the stage for identifying the tame structures that can be used to define observables with such non-analyticities.

\subsection{Non-analytic functions}
There are two major sources for non-analyticity of functions appearing in physics. The first arises from the prototypical example of a function which is smooth but non-analytic, given by
\begin{equation}\label{f-non-an}
f(x) = 
    \begin{cases}
        e^{-1/x} & x > 0\ , \\
        0 & x=0\ .
    \end{cases}
\end{equation}
Every derivative of $f$ vanishes at $x=0$, while $f$ itself is not constant on any neighbourhood of $x=0$. Therefore, the Taylor series
\begin{equation}
    \mathcal{T}f = \sum_{n=0}^{\infty}\frac{x^n}{n!}f^{(n)}(0) \,,
\end{equation}
although it has an infinite radius of convergence, is identically zero. This type of non-analyticity is encountered in the presence of non-perturbative corrections or instantons.

The second source of non-analyticity comes from a rather different reason, namely when the function has a diverging Taylor series with zero radius of convergence. This form of non-analyticity is encountered ubiquitously in partition functions in quantum field theory, due to the factorial growth of the number of Feynman diagrams in perturbative expansions. This is the type of non-analyticity with which we will be concerned in this work.

Let us consider an explicit example of this phenomenon for a zero-dimensional quantum field theory, where an exact computation of the partition function is possible. Consider for instance a $\phi^4$ theory on a point with Euclidean action
\begin{equation}\label{phi^4-point}
    S(\phi; \lambda) = \frac{m^2}{2}\phi^2 + \frac{\lambda}{4!}\phi^4\ ,
\end{equation}
where $\lambda \geq 0$ is a real coupling and $\phi$ is a real scalar field. After the substitutions
    $\phi \rightarrow \sqrt{3/(2\lambda)}\phi$ and $g = 3m^4/(4\lambda)$ we have 
\begin{equation}\label{S0dim}
    S(\phi ;g) = g\phi^2 + \frac{g}{8}\phi^4\ .
\end{equation}
The path integral reduces to an ordinary integral, and the partition function of this theory evaluates to 
\begin{equation}\label{Z(g)}
    Z(g) =\int_{-\infty}^{\infty}\dd\phi \, e^{-S(\phi;g)} =\sqrt{2}e^g K_{1/4}(g)\ .
\end{equation}
Here $K_{1/4}$ is the modified Bessel function, defined by the integral
\begin{equation}\label{K1/4}
    K_{1/4}(z) = \int_{0}^{\infty}\dd t\, \cosh\left(\tfrac{t}{4}\right)e^{-z\cosh(t)} \qquad \text{with}\quad |\arg(z)| < \frac{\pi}{2}\ .
\end{equation}
Although the integral representation is only valid in the right half of the complex plane, $K_{1/4}(z)$ only has one singularity at $z = \infty$, so the above formula holds for all values of the coupling $g \in \mathbb{C}$. The singularity at $\infty$ is essential, and therefore $Z(g)$ is non-analytic in the weak coupling limit $\lambda \rightarrow 0$. Note however hat the limit $\lim_{g\rightarrow \infty}Z(g)$ exists and it is $0$, so that $Z(\lambda)$ is smooth at $\lambda= 0$. The non-analyticity can be seen through the rapid growth of the coefficients in the Taylor expansion of $Z(\lambda)$. We will study this growth in more detail in section \ref{phi4}.

An approach to handle non-analyticity of this type, and hence assign a physical meaning to such divergent perturbative expansions, is to use the technique of Borel resummation. This procedure allows one to to transform the divergent power series into a well-defined function which reproduces the correct expansion coefficients. We will review this technique in the next subsection.

\subsection{Introduction to Borel summability}\label{Borel-summability-section}

Let us briefly introduce some basic facts about Borel summability, mostly following \cite{Sauzin:2014qzt} and \cite{balser1994divergent}. A seminal work about divergent series is \cite{Hardy-book}, while another extensive modern presentation can be found in \cite{LodayRichaud2014DivergentSA}. For more results about multisummability, see \cite{balser1994divergent}.

Consider a divergent (formal) power series 
\begin{equation}\label{formal-power}
    \widetilde{\varphi} = \sum_{n=0}^{\infty} a_n z^{n+1}
\end{equation}
By divergent, it is meant that the power series has zero radius of convergence in the complex plane. The power series $\widetilde{\varphi}$ is said to belong to the Gevrey class $1/p$ if there exist constants $A, B$ such that
 \begin{equation}
     |a_n| \leq A B^n (n!)^p \,,
 \end{equation}
uniformly in $n$. In this case, the Borel operator  $\mathcal{B}_p$ (see e.g. \cite{balser1994divergent}) acting as
\begin{equation}
   \widehat{\varphi}(\zeta)  = \mathcal{B}_p\tilde\varphi = \sum_{n=0}^{\infty} \frac{a_n}{\Gamma(np +1)} \zeta^{n} \ ,
\end{equation}
produces a new series $\widehat \varphi(\zeta)$ with a finite radius of convergence. This operation is known as the (formal) Borel transform of $\widetilde\varphi$. A formal power series $\widetilde\varphi$ of Gevrey class $1/p$ is said to be $p$-summable (or Borel-summable), if its Borel transform $\mathcal{B}_p\widetilde\varphi = \widehat\varphi$ satisfies the following conditions:
    \begin{enumerate}
        \item It admits analytic continuation to a horizontal strip $S_{\delta} = \{ \zeta \in \mathbb{C}:  |\Im{\zeta^{1/p}}| < \delta\}$;
        \item The analytic continuation has exponential size no larger than $1/p$, namely
\begin{equation}\label{exp-size}
    |\widehat\varphi(\zeta)| \leq C \exp\left(c |\zeta|^{1/p}\right)
\end{equation}
for some constants $C$ and $c$, and for sufficiently large $\zeta \in S_\delta$.
    \end{enumerate}
By a slight abuse of notation, we denote the analytic continuation of $\widehat\varphi(\zeta)$ by the same symbol. In case the above conditions are not satisfied, the positive real line $\mathbb{R}^+$ is said to be a Stokes line for $\widetilde\varphi$.

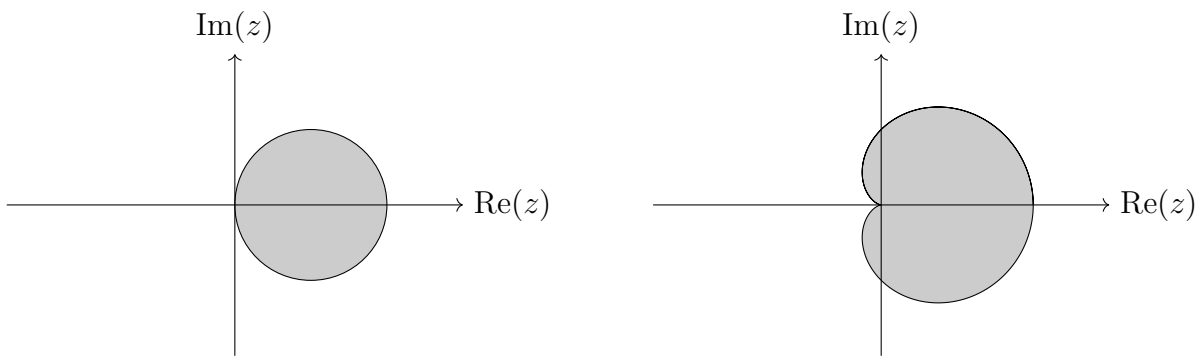
\begin{figure}[h]
    \centering
    \begin{subfigure}[t]{.5\textwidth}
  \centering %
 \begin{tikzpicture}
  \draw[fill=black!20] (1,0) circle [radius=1cm];
      \draw[->] (-3,0) -- (3,0) node[right] {$\Re (z)$};
  \draw[->] (0,-2) -- (0,2) node[above] {$\Im(z)$};
 
\end{tikzpicture}
\end{subfigure}%
\hfill
\begin{subfigure}[t]{.5\textwidth}
  \centering %
 \begin{tikzpicture}
    \draw[smooth,fill=black!20] plot[domain=0:540,samples=200] (\x:{2*cos(\x/2)^2});
  \draw[->] (-3,0) -- (3,0) node[right] {$\text{Re}(z)$};
  \draw[->] (0,-2) -- (0,2) node[above] {$\text{Im}(z)$};
 
\end{tikzpicture}
\end{subfigure}
\caption{Sokal discs for $p=1$ (left) and $p=2$, also known as cardioid (right).}
    \label{fig:Sokal_discs}
\end{figure}

In case these conditions are satisfied, the Laplace operator $ \mathcal{L}_p$ can be applied to $\widehat\varphi$ as
\begin{equation}\label{Laplacetransform}
   \varphi(z)=\mathcal{L}_p\widehat{\varphi} = \frac{1}{p}\int_{0}^{\infty}\dd\zeta\left(\frac{\zeta}{z}\right)^{\frac{1}{p}-1}e^{-\left(\frac{\zeta}{z}\right)^{\frac{1}{p}}}\widehat{\varphi}(\zeta)\ .
\end{equation}
The resulting function $\varphi(z)$ is called the Borel sum of $\widetilde\varphi$, while the procedure described to obtain it is called the Borel resummation. One can then prove that $\varphi(z)$ must be analytic in a domain $D^p_c = \{z \in \mathbb{C} : \textup{Re}(z^{-1/p}) > c\}$, where $c$ is the constant appearing in (\ref{exp-size}). This domain is called a Sokal disc, and it is depicted for $p=1$ and for $p=2$ in figure~\ref{fig:Sokal_discs}. Note that the Sokal disc is open and does not include the origin: although $\varphi(0)$ can be defined as the limit for $z \rightarrow 0$ of $\varphi(z)$, the function thus extended will not be analytic at $z=0$, unless the formal power series (\ref{formal-power}) was actually convergent. The limit $\lim_{z \rightarrow 0}\varphi(z)$ must exist in the Sokal disc if $\widetilde\varphi$ is Borel-summable, and it is actually $0$ if we assume the form (\ref{formal-power}) for the initial formal power series. Intuitively, $z =0$ is the only value for which the power series is actually a function, which evaluates to $0$ at this point. 

The interest in the Borel sum $\varphi(z)$ is due to the Nevanlinna-Sokal theorem \cite{Sokal-improvemnt-on-Watson}.\footnote{For more discussions of this theorem we refer to the appendices of \cite{Magnen_2009,Rivasseau:2023qzm,Ferdinand:2022duk}, and section B of \cite{Lionni:2016ush}.} The theorem guarantees that, if $\varphi(z)$ is the Borel sum of a Borel-summable formal power series $\widetilde\varphi$, of Gevrey class $1/p$, then $\varphi(z)$ admits $\widetilde\varphi$ as asymptotic expansion of order $p$ (indicated by $\varphi(z) \sim_p \widetilde{\varphi}$). More precisely, one has
\begin{equation}\label{asymptotic-expansion}
    \left|\varphi(z) - \sum_{n=0}^{N-1}a_n z^{n+1}\right| \leq A \delta^{-N}(pn!) z^{N+1}
\end{equation}
for every $z$ in the Sokal disc $D^p_c$.\footnote{Note that we can always substitute $(pn!)$ with $(n!)^p$ with a due rescaling of the constants $A$ and $\delta$ without altering the Gevrey class of the asymptotic expansion.} Observe how the exponential growth of the upper bound depends inversely on the width of the sector in which the Borel transform admits analytic continuation with exponential size at most $1/p$.

Most importantly, the theorem also holds in reverse. That is, the existence of a function $\varphi(z)$ which is analytic on the Sokal disc $D^p_c$ and admits a uniform asymptotic expansion as (\ref{asymptotic-expansion}) on $D^p_c$ suffices to conclude that the formal power series $\widetilde\varphi$ is Borel-summable, and that $\varphi(z)$ is its unique Borel sum. The combination of the two statements of the Nevanlinna-Sokal theorem ensures that there is a one-to-one correspondence between the algebra of functions holomorphic on a Sokal disk and the algebra of their asymptotic expansions. In other words, the Taylor map on this algebra of functions is injective: this feature is known as quasi-analyticity.

The Nevanlinna-Sokal theorem is at the core of all the results that will follow: in particular, we will use the converse statement in section \ref{poly-potential} to prove the Borel summability of a path integral with a polynomial interaction.

\section{Tame geometry and the o-minimal structure $\bbR_{\mathscr{G}}$}\label{o-minimality-section}
The goal of this work is to demonstrate that non-analytic partition functions and correlation functions, which can be examined as Borel sums of their asymptotic perturbative expansions, belong to a specific class of functions known as tame functions. The special defining feature of these functions is that they are definable in an o-minimal structure. In this section, we briefly review the necessary material from tame geometry to make this statement precise. Specifically, we introduce the o-minimal structure called $\Rg$, which allows for defining Borel-resummed functions.

\subsection{Basics on o-minimal structures and first examples}

O-minimal structures are geometrical objects which reflect the properties of first-order languages in logic. We refer to \cite{MR1633348} for an extensive introduction, and provide here a brief definition, which can be also found in \cite{Douglas:2022ynw,Douglas:2023fcg,Grimm:2023xqy}.

An \textit{o-minimal structure} is a collection $\mathcal{S} = (S_n)_{n \in \mathbb{N}}$, where  every $S_n$ is a family of subsets $\mathbb{R}^n$, satisfying the following requirements:
    \begin{enumerate}
        \item For every $n \in \mathbb{N}$, $S_n$ is a Boolean algebra: namely, it is closed under intersections, unions and complements. That is, for every $A, B \in S_n$, one has that $A \cup B$, $A\cap B$, $A^c$ all belong to $S_n$.
        \item $\mathcal{S}$ is closed under Cartesian products: namely, for every $A \in S_n$, and $B \in S_m$, $A \times B$ and  $B \times A$ belong to $S_{n+m}$. Moreover, $\mathcal{S}$ is closed under projections: namely, $\pi (A) \in S_{n-1}$, where $\pi : S_n \to S_{n-1}$ is the projection onto the first $n-1$ coordinates.
        \item $\mathcal{S}$ contains all the zero sets of all polynomials in $n$ real variables.
        \item If $A \in S_1$, it can be written as a finite union of points and intervals.
    \end{enumerate}

The first three axioms define what is called a structure. The fourth axiom is a tameness axiom, which ensures that the objects in the structure have a built-in notion of finiteness. A set $A$ which belongs to $S_n$ for some $n$, is said to be $\mathcal{S}$-definable. More generally, we call $A$ definable, or tame, when the explicit the underlying o-minimal structure is clear from the context.

The notion of tameness extends to functions by viewing their graphs as definable sets. Let $f : A \to \mathbb{R}$ be a function, where $A \in S_n$ is $\mathcal{S}$-definable for some o-minimal structure $\mathcal{S}$. Then we shall say that $f$ is $\mathcal{S}$-definable if its graph
     \begin{equation}
         \Gamma(f) = \{(x,y) \in A \times \mathbb{R} \; : y = f(x) \}
     \end{equation}
 is a $\mathcal{S}$-definable set. Note that sums, products and compositions of $\mathcal{S}$-definable functions are still $\mathcal{S}$-definable \cite{MR1633348}.

Let us now discuss some examples of o-minimal structures. The smallest o-minimal structure, i.e.~the smallest collection of sets which is compatible with the above axioms, is called $\bbR_{\rm alg}$ and consists of semi-algebraic sets. These are sets which can be written as finite unions and intersections of sets defined by polynomial equalities and inequalities, of the form
\begin{equation}
     A= \{ x \in \mathbb{R}^n : P(x) = 0, Q(x)>0\},
\end{equation}
where $P, Q_1,\ldots,Q_r$ are polynomials of  the ring $\mathbb{R}[x_1,\ldots,x_n]$.

More general o-minimal structures are typically obtained by specifying a set of functions $\cF$ and considering the structure $\bbR_\cF$ generated by $\cF$, by which we mean the smallest o-minimal structure containing the graphs of all the functions in $\cF$. The collection $\cF$ may consist of a single function, such as the real exponential function which generates the structure $\bbR_\text{exp}$, or an entire class of functions, such as the restricted analytic functions which generate the structure $\bbR_\text{an}$. The latter class of functions are defined by restrictions of analytic functions to the cube $[-1,1]^n$. This restriction to a compact domain is crucial for ensuring that these functions have sufficiently tame behavior and therefore that $\bbR_{\rm an}$ satisfies the o-minimality axiom.

Another o-minimal structure which we highlight is $\bbR_{\rm Pfaff}$, which is generated by Pfaffian functions. Given an open set $U \subseteq \mathbb{R}^m$, a Pfaffian chain of length $n$ is defined to be a triangular system of $n$ coupled differential equations in $m$ variables, solved by $n$ functions $\zeta_1, \ldots, \zeta_n : U \rightarrow \mathbb{R}$ of the form
$\partial\zeta_i/\partial x_j = F_{ij}(x_1,\ldots, x_m, \zeta_1, \ldots,\zeta_{i})$,
where $F_{ij}$ are a polynomials in both the coordinates $x_j$ and the functions $\zeta_i$. The structure $\bbR_{\rm Pfaff}$ is generated by all solutions to such differential equations. Its primary interest comes from the fact that it admits a precise notion of complexity, which was applied to physical settings in \cite{Grimm:2023xqy,Grimm:2024mbw}.

In the previous section, it has been argued why partition and correlation functions are non-analytic in the weak-coupling limit. Therefore, in order to prove that they are tame, we need an o-minimal structure which hosts non-analytic functions. The collection $\mathcal{F}$ of the structures listed above, $\bbR_{\textup{an}}$ and $\mathbb{R}_{\textup{Pfaff}}$, only comprise analytic functions. This does not mean that all the definable functions of these structures are analytic: for instance, the function $f(x)$ defined by (\ref{f-non-an}) can be shown to be definable in $\bbR_{\rm exp}$ and $\bbR_{\textup{Pfaff}}$ by composing elementary functions. This procedure is however not sufficient to understand the tameness of non-analytic functions in greater generality. For example, the partition $Z(\lambda)$ of the $\phi^4$ theory on a point discussed in the previous section was only shown to be definable in $\bbR_{\rm Pfaff}$ away from the weak-coupling limit \cite{Grimm:2023xqy}. In the following we will describe a larger o-minimal structure, called $\Rg$, which contains non-analyticity in a more refined way, allowing us to extend this analysis.

\subsection{The tame structure of Gevrey functions $\mathbb{R}_{\mathscr{G}}$}\label{Rg-section}

The o-minimal structure $\Rg$ was introduced in \cite{DRIES_SPEISSEGGER_2000}. In this structure, the collection $\mathcal{F}$ comprises all the Gevrey functions in any number of dimensions. Moreover, $\mathcal{F}$ can also include the exponential $\exp : \bbR \rightarrow \bbR$, in which case the structure is named $\Rge$. The complete definition of $\mathcal{F}$ is quite technical, and we refer to Appendix \ref{Rg-many-variables} for a brief review and to \cite{DRIES_SPEISSEGGER_2000} for the details. For our purposes though, we will be concerned with functions of $\mathcal{F}$ of only one variable, which admit a much simpler description. Consider the holomorphic functions $f(z)$ on a sector
\begin{equation}
    S(R,\phi, p) = \{ z \in \mathbb{C} \; :\; 0< |z| < R , |\textup{arg}(z)| < p \phi\}
\end{equation}
where $R \in (0, \infty)$, $0< p \leq 1$ and, most importantly, $\phi \in (\frac{\pi}{2}, \pi)$, satisfying the following conditions:
\begin{enumerate}
    \item the limit $\textup{lim}_{z \rightarrow 0}f^{(n)}(z)$ when taken in $ S(R,\phi, p)$ for every $n \in \mathbb{N}$;
    \item the function $f$ satisfies the Gevrey condition: there exist constants $A, B$ depending on $f$, such that
\begin{equation}\label{GevreyvdD}
    \left|\frac{f^{(n)}(z)}{n!}\right| \leq A B^n (n!)^{p}
\end{equation}
\end{enumerate}
 The first condition ensures that $f(z)$ admits an asymptotic expansion on $S(R, \phi, p)$, and the second ensures that the asymptotic expansion is of Gevrey class $1/p$  \cite{LodayRichaud2014DivergentSA}. Moreover, the condition $\phi \in (\frac{\pi}{2}, \pi)$ ensures that the Nevanlinna-Sokal theorem holds, and therefore $f(z)$ is equal to the Borel sum of its asymptotic expansion. Although the sector $S(R, \phi, p)$ is larger than a Sokal disc $D^p_R$, it can be understood in light of the results of \cite{balser1994divergent}\footnote{The relevant result is Lemma 1 of section 3.2 in \cite{balser1994divergent}.} that the above conditions are not more restrictive: roughly, the analyticity domain where the Gevrey condition holds can be extended beyond the Sokal disc by rotating the line of integration of the Laplace transform by an angle $\varepsilon$, which is always possible for $\varepsilon$ small enough (provided that Borel summability holds).

 Since the limit for $z\rightarrow 0$ must exist, we are enabled to extend every such function $f$ to $S \cup \{0\}$ by setting $f(0) := \textup{lim}_{z \rightarrow 0}f(z)$. The tame functions of one variable $\tilde f \in \mathcal{F}$ are then defined on the interval $[0, R]$, with
 \begin{eqn}
     \begin{cases}
         \tilde f(x) = f(0) & x =0;\\
         \tilde f(x) = f(x) & x \in (0,R]
     \end{cases}
 \end{eqn}
This means that the tame functions $\tilde f $ are Borel sums, non-analytic in the weak-coupling limit $x=0$. Moreover, observe how, when $p> 1$, by a simple change of variable $z \rightarrow z^p$ it is possible to turn the Gevrey class of a formal power series from  $1/p$ to  $1$; it is also easy to check that Borel summability is preserved under this operation (see \cite{LodayRichaud2014DivergentSA}, Section 2.3. for details). Thus, all the Borel sums of Borel-summable asymptotic series, defined on a domain which includes the origin, are definable in $\Rg$.

Before moving to the next section, let us briefly summarize our strategy to prove the tameness of partition functions and correlation functions in terms of the coupling $\lambda$. The first step is to perform the perturbative expansion, obtaining an asymptotic series $\widetilde \varphi$. Next, we will compute the Borel transform $\widehat\varphi = \mathcal{B}_p\tilde\varphi$ and then use the fact that if $\widehat\varphi(\zeta)$ has no poles on $\bbR^+$ and exponential size less than $1/p$, then the series $\widetilde\varphi$ is $p$-summable along $\bbR^+$. Consequently, the resulting Borel sum $\varphi(\lambda)$ is definable in the structure $\Rg$.

\section{Tame partition functions in 0-dimensional QFTs}\label{0-dim}

In this section we discuss partition functions of simple interacting quantum field theories on a point-like space-time and prove that they are tame functions of $\Rg$. After taking $\phi^4$ theory as our starting point, we continue to discuss real and complex scalar theories with a higher order monomial interaction term. We then proceed with a discussion of matrix theories. Finally, we prove the tameness of partition functions for a scalar field theory with a general polynomial interaction. Throughout this section, we use the symbol $\widetilde\varphi$ for asymptotic series, $\widehat\varphi$ for the corresponding Borel transforms, and $\varphi$ for the Borel-resummed functions. 

\subsection{Real $\phi^4$ theory on a point}\label{phi4}
To begin with, we return to our initial example \eqref{phi^4-point} and provide a new perspective. Let us proceed perturbatively, expanding the quartic interaction and finding the asymptotic expansion
\beq\label{Z-expansion}
    Z(g) = \int_{-\infty}^{\infty}\dd\phi \,e^{-g\phi^2}\sum_{n=0}^{\infty}\frac{(-1)^n}{n!}\left(\frac{g\phi^4}{8}\right)^n 
    \quad \sim_1\quad \sum_{n=0}^{\infty}\frac{(-1)^n}{n!}\left(\frac{g}{8}\right)^n \int_{-\infty}^{\infty}\dd\phi\, e^{-g\phi^2}\phi^{4n}.
\eeq
It is of great importance to highlight that in passing from the first to the second infinite sum we have changed the order of the the integral with the sum, and hence we can no longer claim that the right-hand expansion is equal to the left-hand expansion. Instead, we can only claim that it is its asymptotic expansion and therefore use the symbol $\sim$. The Gaussian integral can be expressed as a Gamma function, yielding the formal power series
    \begin{equation}\label{Z-perturb2}
   Z(g) \sim_1 \sum_{n=0}^{\infty}\frac{(-1)^n}{n!}\frac{1}{8^ng^{n+1/2}}\Gamma\left(2n +\frac{1}{2}\right)=:\sqrt{\pi g}\widetilde{\varphi}(g)\, .
\end{equation}
Here we have extracted a factor $\sqrt{\pi g}$ for convenience. Setting $\widetilde\varphi = \sum_{n=0}^{\infty}a_ng^{-n-1}$ and recalling that  $\Gamma(n/2) = \sqrt{\pi}2^{-(n-1)/2}(n-2)!!$ for any odd $n$, one finds that the coefficients $a_n$ are given by
\begin{equation}
    a_n = \frac{(-1)^n}{n!}\frac{(4n-1)!!}{32^n}\,.
\end{equation}
These coefficients diverge factorially and therefore the Gevrey class of $\widetilde\varphi$ is $1$. Observe how $(4n-1)!!$ is the number of Wick contractions among $4n$ fields under Gaussian integration, i.e. the number of Feynman diagrams; it can now be seen explicitly that they their number indeed grows faster than $n!$, as claimed before. The Borel transform of $\widetilde{\varphi}$ is\footnote{We are now working in the large $g$ limit rather than the small $\lambda$ limit, but this does not change the previous discussion about Borel summability.}
\begin{equation}
        \widehat{\varphi}(\zeta) = \sum_{n = 0}^{\infty}\frac{(-1)^n}{32^n}\frac{(4n-1)!!}{(n!)^2}\zeta^n = \frac{2}{\pi}\frac{1}{(1+\zeta/2)^{1/4}} K\Big(\tfrac{1}{2} -\tfrac{1}{2\sqrt{1+\zeta/2}}\Big)\ ,
\end{equation}
where $K(z)$ is the complete elliptic integral of the first kind. The only pole of $\widehat\varphi(\zeta)$ is $\zeta = -2$, and the reason that it lies on the negative real axis is that the coefficients of the asymptotic series $\widetilde\varphi$ has alternating signs. Furthermore, it can be checked that in the large $\zeta$ limit, $\widehat{\varphi}(\zeta)$ is decaying. Since $\widehat\varphi(\zeta)$ has no poles on $\bbR^+$, we conclude that $\widetilde\varphi$ is $1$-summable, and the asymptotic expansion is of order $1$, i.e.~$Z(g) \sim_1 \sqrt{\pi g}\widetilde\varphi(g)$ holds uniformly on (every closed subsector of) $\bbC \backslash \bbR^-$. The Borel sum 
\begin{equation}
    \varphi(g) = \int_0^{\infty}\dd\zeta e^{-g\zeta}\widehat\varphi(\zeta)
\end{equation}
is then definable in the structure $\Rg$, from which we conclude that $Z(g) = \sqrt{\pi g}\varphi(g)$ is also definable in $\Rg$. The resummed function is indeed given by $\sqrt{\pi g}\varphi(g) = \sqrt{2}e^g K_{1/4}(g)$, as reviewed in section \ref{sec:nonanal-Borel}.  

We remark that Borel summability of this simple example is also proved in \cite{Rivasseau_2009} by means of the so-called loop vertex expansion and exploiting the Nevanlinna-Sokal theorem, rather than computing the Borel transform of the perturbative power series explicitly. 

\subsection{Real  $\phi^{2p}$  theory on a point}\label{phi-2p}

The results of subsection \ref{phi4} can be extended to a more general potential. Let us consider the action
\begin{equation}\label{2paction-real}
    S(\phi; \lambda) = \frac{m^2}{2}\phi^2 + \lambda \phi^{2p}
\end{equation}
where $p>1$ is an integer. Upon setting $y^2 = \frac{m^2}{2}\phi^2$  and redefining the coupling with $ \lambda \left(\frac{2}{m^2}\right)^{p}\rightarrow \lambda$, the partition function reads
\begin{equation}
    Z(\lambda) = 
    \frac{\sqrt{2}}{m} \int_{-\infty}^{\infty} \dd y \, e^{-y^2-\lambda y^{2p}}\ .
\end{equation}
Again proceeding perturbatively, in analogy to the previous case, we find the asymptotic expansion
\begin{equation} \label{def-tildephi}
        Z(\lambda) \sim_{p-1} \frac{\sqrt{2}}{m}\sum_{n=0}^{\infty}\frac{(-1)^n}{n!}\lambda^n\Gamma\left( np + \frac{1}{2}\right) =:\frac{\sqrt{2}}{\lambda m}\widetilde\varphi(\lambda)
\end{equation}
where now $\widetilde\varphi(\lambda)$ is a formal power series of Gevrey class $1/(p-1)$.

To check that $Z(\lambda)$ is definable in $\Rg$, we compute the Borel transform of \eqref{def-tildephi}. As detailed in appendix \ref{Borel-explicit}, the Borel transform of $\widetilde\varphi(\lambda)$, $\mathcal{B}_{p-1}\widetilde\varphi = \widehat\varphi(\zeta)$ can be computed explicitly in terms of the generalized hypergeometric function ${}_a F_b \left[{\vec{a} \atop \vec{b}} \Big| z\right ]$. Explicitly, we find
\begin{equation}\label{hyper-sol}
    \widehat{\varphi}(\zeta) = \sqrt{\pi}\,_aF_b\left[ {\vec{a} \atop \vec{b}} \bigg | -\frac{p^p}{(p-1)^{p-1}}\zeta \right]\ ,
\end{equation}
where $a = |\vec{a}|$, $b = |\vec{b}|$ with $a=p$, $b = p-1$ and
\begin{equation}\label{abvectors}
    \vec{a} = \frac{1}{2p}\left(1, 3,\ldots, 2p-1\right) \ , \qquad 
    \vec{b} = \frac{1}{p-1}\left(1, 2,\ldots,  p-1\right)\ .
\end{equation}

The poles of the hypergeometric function ${}_a F_b \left[{\vec{a} \atop \vec{b}} \Big| z\right ]$ lie at $z =1$ whenever $a = b+1$ \cite{Douglas:2022ynw}. Hence the unique pole in the Borel plane of $\widehat\varphi(\zeta)$ lies at $\zeta = -(p-1)^{p-1}/p^p$. Therefore it is analytic in a horizontal strip $S_\delta$ and it can also be argued that it has no exponential growth thereon. The asymptotic expansion $\widetilde\varphi$ is then $(p-1)$-summable and its Borel sum $\varphi(\lambda) = \mathcal{L}_{p-1}\widehat\varphi$ is definable in $\Rg$ . We finally conclude that the partition function for the action (\ref{2paction-real}), $Z(\lambda) = \frac{\sqrt{2}}{m\lambda}\varphi(\lambda)$, is definable in $\Rg$ for any integer $p$.

We remark that this kind of partition function was studied extensively in \cite{Fauvet_2020}. There, the resurgence properties of the partition functions, and in particular the location of the poles in the Borel plane, were determined from the differential equation obeyed by the partition function by means of the Newton polygon. The order of this differential equation is $p$. In particular, only for $p \leq 2$ the ODE is of first or second order and can then be recast into a Pfaffian system via the associated Riccati equation, implying that the partition function is definable in $\bbR_{\textup{Pfaff}}$ (see \cite{Grimm:2023xqy} for examples). For larger $p$, we can no longer rely on Pfaffian systems and we need to resort to $\Rg$ to prove that $Z(\lambda)$ is tame.

The correlation functions in this theory can be shown to be Borel-summable in a similar fashion. The $j$-th moment $G_j(\lambda)$ is given by
\begin{equation}
    G_j(\lambda) = \int_{-\infty}^{\infty}\dd\phi \, \phi^{2j}e^{-\phi^2-\lambda\phi^{2p}}.
\end{equation}
As before, expanding the exponential yields the formal power series of Gevrey class $1/(p-1)$
\begin{equation}
    \frac{1}{\lambda}\widetilde{\varphi}_j=\sum_{n=0}^{\infty}\frac{(-\lambda)^n}{n!}\Gamma\left(np+j+\tfrac{1}{2}\right).
\end{equation}
The Borel transform is then computed to be
\begin{equation}\label{hyper-sol-j}
    \widehat{\varphi}_j(\zeta) := \mathcal{B}_{p-1}\widetilde\varphi_j = \Gamma\left(\tfrac{1}{2}+j\right) \;_aF_b\left[ {\vec{a} \atop \vec{b}} \Bigg | -\frac{p^p}{(p-1)^{p-1}}\zeta \right]\,,
\end{equation}
where again $a = |\vec{a}|$, $b = |\vec{b}|$ with $a=p$, $b = p-1$ and
\begin{equation}\label{abvectors}
    \vec{a} = \frac{1}{2p}\left(1+2j, 3+2j,\ldots, 2p-1+2j\right), \qquad
    \vec{b} = \frac{1}{p-1}\left(1, 2,\ldots,  p-1\right). 
\end{equation}
The remainder of the proof is similar to the previous case and is reported in appendix \ref{Borel-explicit}. The location of the pole in the Borel plane is independent of $j$, and Borel summability follows from the properties of the hypergeometric function. Thus, the function $\varphi_j(\lambda) = \mathcal{L}_{p-1}\widehat\varphi_j$ is definable in $\Rg$ and so is the moment $G_j(\lambda) =\tfrac{1}{\lambda }\varphi_j$. Consequently, the correlation function
\begin{equation}
     \langle \phi^{2j}\rangle = \frac{\int_{-\infty}^\infty \dd\phi \,\phi^{2j} e^{-\phi^2-\lambda\phi^{2p}}}{\int_{-\infty}^\infty \dd\phi e^{-\phi^2-\lambda\phi^{2p}}}
\end{equation}
is also definable in $\Rg$, as it is the ratio between two definable functions of the same structure.

\subsection{Complex $(\phi\bar\phi)^{p}$ theory on a point }
Let us now consider again a monomial potential, but for a complex field $\phi$. We will once more proceed perturbatively; let us remark though, that this case was has been already studied in \cite{Rivasseau_2017} using constructive techniques. Constructive field theory (see e.g the introduction of \cite{Rivasseau_2007} and \cite{Rivasseau_2009} for a brief pedagogical review) provides an alternative to perturbative field theory, avoiding the issue of divergence. While perturbative field theory produces a divergent power series in powers of the coupling $\lambda$, whose coefficients are computed by sums over Feynman diagrams, constructive field theory forgoes this common notion and replaces it by a convergent sum over \textit{forests} or \textit{spanning trees}.

The deep insight of constructive field theory is that the full knowledge of loops makes the sums over Feynman diagrams diverge, whereas spanning trees, although encoding the same amount of information, have no loops and thus their sum does not diverge. We refer to \cite{abdesselam1995trees} for a detailed discussion and proofs; to \cite{Rivasseau_2013} for the relationship between trees and connected Feynman diagrams.

Before relying on these more advanced tools, let us proceed perturbatively once more. The partition function for a complex field in $0$ dimensions is
\begin{equation}
    Z_p(\lambda) = \int \tfrac{\dd\phi \, \dd\bar{\phi}}{2\pi i}e^{-\phi\bar\phi-\lambda(\phi\bar\phi)^p}.
\end{equation}
The integral measure $\tfrac{\dd \phi \dd \bar\phi}{2\pi i} $ can be rewritten in polar coordinates as $\frac{2ir}{2\pi i}\dd\theta \wedge \dd r$. Hence the partition function takes the form
\begin{equation}
   Z_p(\lambda) = \int_0^{\infty}\dd x \,e^{-x-\lambda x^{p}}.
\end{equation}
We can now find the its asymptotic expansion in power series in $\lambda$ given by
\begin{equation}
    \frac{1}{\lambda}\tilde\varphi(\lambda) := \sum_{n=0}^{\infty} \frac{(-1)^n\lambda^n}{n!}\int_{0}^{\infty}\dd x\, e^{-x}x^{pn} = \frac{1}{\lambda}\sum_{n=0}^{\infty} (-1)^n\lambda^{n+1} \frac{(pn)!}{n!}\, ,
\end{equation}
which is of Gevrey class $1/(p-1)$ and thus $Z_p(\lambda) \sim_{p-1} \frac{1}{\lambda}\widetilde\varphi(\lambda)$. The Borel transform of the above sum is then
\begin{equation}\label{Borel-trans- cplx}
    \widehat{\varphi}(\zeta) = \sum_{n=0}^{\infty}(-1)^n\zeta^n\frac{pn!}{n!((p-1)n)!} = \sum_{n=0}^{\infty}(-1)^n\zeta^n\binom{pn}{n}\, ,
\end{equation}
which is exactly the function $F_p(-z)$ in the notations of \cite{Rivasseau_2017}. As observed therein, the binomial coefficient $\binom{pn}{n}$ is closely related to the $n$-th Fuss-Catalan number
\begin{equation}
    C^{(p)}_n = \frac{1}{pn+1}\binom{pn+1}{n} = \frac{1}{(p-1)n+1}\binom{pn}{n}\, .
\end{equation}
Fuss-Catalan numbers are associated with the generating function $T_p(z) = \sum_{n=0}^{\infty}C^{(p)}_nz^n$, which satisfies the algebraic relation $zT_p^p(z)+1 = T_p(z)$. It is then easy to prove that
\begin{equation}
    \widehat\varphi(\zeta) = \frac{1}{1+p\zeta T_p^{p-1}(-\zeta)}\,.
\end{equation}
According to Theorem III.1 in \cite{Rivasseau_2017}, the radius of convergence of the Borel transform $\widehat\varphi(\zeta)$ is the same of that of the Fuss-Catalan generating function $T_p(z)$, which is $R_p = \tfrac{(p-1)^{(p-1)}}{p^p}$. As it might have been expected, this is the same radius of convergence of the Borel transform in the case of the real field dealt with in the previous section. Again according to Theorem III.1, $ \widehat\varphi(\zeta)$ is holomorphic on $\mathbb{C}\backslash[-R_p, -\infty]$ and its growth for $\Re(\zeta) \rightarrow \infty$ is polynomially bounded: $Z_p(\lambda)$ is therefore $(p-1)$-summable and is equal to the Borel sum of its asymptotic expansion.

As in the previous cases, Borel summability is owed essentially to the alternate signs in the Borel transform  (\ref{Borel-trans- cplx}), which cause the pole to lie on the negative real axis. We can then conclude that $Z_p(\lambda)$ is definable in $\Rg$ for all  positive integers $p$.

\subsection{Complex matrix theory on a point}
    
We now wish argue for the tameness in $\Rg$ of the partition functions of more general models on a point, for instance matrix models. In this subsection we will rely on Borel-summability results from the literature obtained using constructive field theory.

First, a matrix model with a quartic interaction was studied constructively in \cite{Rivasseau_2007} by means of the Loop Vertex Expansion, extending the method portrayed in \cite{Rivasseau_2017} to a complex $N\times N$ matrix field $M$. The partition function in question is
    \begin{equation}
        Z(\lambda, N) =  %\pi^{-N^2}%
         \int \dd M \dd M^{\dagger}  \, e^{-\tfrac{1}{2}\Tr{M^\dag M}-\frac{\lambda}{N}\Tr{M^\dag M M^\dag M}} \,.
    \end{equation}
  
Borel summability is then proved by performing first a Hubbard-Stratonovich transformation, thus trading a quartic interaction for a cubic one; then by integrating out the original field $M$, finding thus the loop vertices; finally, by applying the BKAR forest formula (see \cite{abdesselam1995trees} for an exhaustive review, \cite{Rivasseau_2009} for a more concise explanation). Theorem 3.1 of \cite{Rivasseau_2007} guarantees Borel summability for $Z(\lambda, N)$  (more specifically, $1$-summability) uniformly in $N$. Therefore, we are able to conclude that the partition function for this model is also definable in $\Rg$.

The previous result was extended to higher order interactions in \cite{Lionni:2016ush}; the results were further improved in \cite{Krajewski:2017thd, Krajewski:2019tsi} employing the Loop Vertex Representation
and the Taylor BKAR forest formula.  The model considered has a partition function 
\begin{equation}
    Z(\lambda, N) = \int \dd M \dd M^{\dagger} \exp\left(-\textup{Tr}(MM^{\dagger})-\tfrac{\lambda}{N^{p-1}}\textup{Tr}(MM^{\dagger})^p\right) \,.
\end{equation}
This partition function was proved to be $(p-1)$-summable in Theorem 3 of \cite{Lionni:2016ush}. More precisely, it was proved to be analytic and to admit Gevrey bounds on a Sokal disc
\begin{equation}
    D^{p-1}_{\rho(p,N)} := \{\lambda \in \bbC: \Re \lambda^{-1/(p-1)} >  \rho(p,N) \}
\end{equation}
where $\rho(p,N) = \rho_pN^{1+2/(p-1)}$. This disc shrinks as $N$ increases, but it was proved in  \cite{Krajewski:2017thd, Krajewski:2019tsi} that the analyticity domain can be extended to a domain $P(\epsilon) := \{ \lambda \in \mathbb{C} : 0 < |\lambda| < \eta \;, |\arg{\lambda}| < \pi -\epsilon\}$ which is uniform in $N$. 
We conclude that $Z(\lambda, N)$, once restricted to an interval including the origin, is definable in $\Rg$, for every $N$. We remark that, because $\bbR_{\textup{an}}$ is contained in $\Rg$, given a definable function of $\Rg$ defined on an interval $[0, r_1]$, its analytic continuation to an interval $[0,r_2]$ remains definable in $\Rg$ as long as $r_1$ is larger than $0$ and $r_2$ is finite.

Another recent improvement on this result is provided in \cite{Rivasseau:2023qzm}. By considering a partition function with sources 
\begin{equation}
    Z(\lambda, N, J,  J^{\dagger}) = \int \dd M \dd M^{\dagger} \exp\left(- \textup{Tr}(MM^{\dagger})- \tfrac{\lambda}{N^{p-1}}\textup{Tr}(MM^{\dagger})^p + \sqrt{N} \textup{Tr}(JM^{\dagger}) + \sqrt{N} \textup{Tr}( MJ^{\dagger})\right)
\end{equation}
we can compute the following $2k$-point function
\begin{equation}
    G^k(\lambda, N):= \left[\frac{\partial^2}{\partial J^*_{a_1, b_1} \partial J_{c_1, d_1}}\cdots \frac{\partial^2}{\partial J^*_{a_k, b_k} \partial J_{c_k, d_k}}\right]\log Z(\lambda, N, J, J^{\dagger}) \Bigg|_{J = J^{\dagger} = 0}
\end{equation}
where $J^*_{a_i, b_i} = J^{\dagger}_{b_i a_i}$.  The result of \cite{Rivasseau:2023qzm} is then that these functions are analytic in the cardioid domain (recall Figure \ref{fig:Sokal_discs})
\begin{equation}\label{cardiod domain}
    \mathcal{C} = \left\{\lambda \in \mathbb{C} : |\lambda| < \frac{1}{2(p-1)} \cos^{p-1}\left(\frac{\arg{\lambda}}{p-1}\right) \right\}
\end{equation}
uniformly in $N$, provided that $||{JJ^{\dagger}}|| < \epsilon$ for some $\epsilon$ depending on $\lambda$. 
Moreover, the same theorem guarantees that the $2k$-point functions admit an asymptotic expansion of Gevrey class $1/(p-1)$. Note that the cardioid domain is in fact a Sokal disc $D^{p-1}_c$ with $c= (2(p-1))^{p-1}$, which takes on the shape of a usual disc for $p=1$. The theorem then ensures Borel summability. Consequently, we conclude that the $2k$-point functions are also definable in $\Rg$ as functions of the coupling $\lambda$.

\subsection{Scalar with polynomial potential}\label{poly-potential}
So far we have considered theories with a single monomial interaction term. In this section we consider again a scalar theory on a point, but with a general polynomial potential. In this case, the explicit computation of the Borel transform is not possible, and we must rely on the Nevanlinna-Sokal theorem to prove Borel summability. Consider the partition function for a 0-dimensional, scalar bosonic QFT
\begin{equation}\label{Z-def-poly}
    Z(\lambda) = \int_{-\infty}^{\infty}\dd\phi \,e^{-\phi^2 -\lambda V(\phi)}\ ,
\end{equation}
where we assume that $V(\phi)$ is a polynomial of degree $2p$, given by 
\begin{equation}
    V(\phi) = a_{2p}\phi^{2p}+ a_{2p-1}\phi^{2p-1}+\ldots+a_3\phi^3\ .
\end{equation}
We assume that the leading coefficient satisfies $a_{2p} > 0$ so that the integral is well-defined for a positive coupling $\lambda$. Moreover, under our assumptions, $V(0) =0$. Then $Z(\lambda)$ can be shown to be $(p-1)$-summable by using the Nevanlinna-Sokal theorem. This requires two steps: first, finding an analyticity domain for $Z(\lambda)$ which contains a Sokal disc $D^{p-1}_0$; second, proving that it is asymptotic to a $1/(p-1)$-Gevrey formal power series on that domain. 

Before showing the proof, let us observe that we can reasonably expect $Z(\lambda)$ to inherit $(p-1)$-summability from the partition function for the monomial potential $V(\phi) = \phi^{2p}$ discussed in section \ref{phi-2p}. Indeed, the formal power series
\begin{equation}
    \widetilde\varphi:= \sum_{n=0}^{\infty}\frac{(-\lambda)^n}{n!}\int_{-\infty}^{\infty}\dd\phi \,e^{-\phi^2}V^{n}(\phi)
\end{equation}
has alternating signs for $n$ larger than a certain $n_{\rm max}$, because eventually the Gaussian integration of leading term $(a_{2p}\phi^{2p})^n$, which grows  as $\Gamma(np+1)$, will dominate the other terms completely. This feature pushes the pole in the Borel plane on the negative real axis $\bbR^-$, ensuring Borel summability.

\subsubsection{Analyticity domain}
The integral representation (\ref{Z-def-poly}) is analytic on $\{\lambda \in \bbC : \Re\lambda > 0\}$, but this domain can be extended by means of a change of variable. We let $\phi^2 = \lambda^{-(q-1)/(qp)}x^2$, for a general positive integer $q$. Since we are eventually interested in a real coupling $\lambda$, we make the change of variable thinking of $\lambda$ as being real. Subsequently, we then perform the analytic continuation to complex values. Writing $V(\phi) = v(\phi) + \phi^{2p}$ and putting $a_{2p} =1$ for simplicity, we then have
\begin{equation}\label{Z-change-of-variable}
     Z(\lambda) = \int_{-\infty}^{\infty}\dd x\,\lambda^{-\frac{q-1}{2qp}}\exp\left(-\lambda^{-\frac{q-1}{qp}}x^2-\lambda^{\frac{1}{q}}x^{2p}-\lambda v\left( \lambda^{-\frac{q-1}{qp}}x\right)\right).
\end{equation}
$Z(\lambda)$ is now analytic on  the Sokal disc $D_{0}^q = \{ \lambda \in \bbC: \Re \lambda^{\frac{1}{q}} > 0\}$, namely on $\{\lambda \in \bbC:|\arg \lambda | < q \frac{\pi}{2}\}$.\footnote{To be more precise, as the $Z(\lambda)$ has a branch cut on $\bbR^-$, Sokal discs with $D^q_0$ with $q>2$ should be viewed as subsets of the Riemann surface of the logarithm (see e.g \cite{Sauzin:2014qzt}).} The analyticity domain can then be extended indefinitely, by letting $q$ be large enough. However, we do not merely seek an analyticity domain, but an analyticity domain whereon the function $Z(\lambda)$ admit an asymptotic expansion. As stated in \cite{LodayRichaud2014DivergentSA}, proposition 2.2.11, this is true if and only if the limit for $|\lambda| \rightarrow 0$  of $Z(\lambda)$ and all its derivatives exists on every direction of $D_0^q$. If we change variable again to $y^2 = |\lambda|^{-(q-1)/(qp)}x^2$ we clearly find
\begin{equation}
    Z(\lambda) = \int_{-\infty}^\infty \dd y \,e^{-i \theta \frac{q-1}{2qp}}\exp\left(-y^2e^{-i \theta \frac{q-1}{qp}} - |\lambda| \left(e^{-i\frac{\theta}{q}}y^{2p} + v\left(e^{-i\frac{q-1}{2qp}\theta}y\right)\right)\right)
\end{equation}
where $\lambda = |\lambda| e^{i\theta}$. From this we can infer that the limit for $|\lambda| \rightarrow 0$, with $\lambda \in D^{q}_0$, will only exist if  $q \leq p+1$. The largest domain of analyticity in which $Z(\lambda)$ admits an asymptotic expansion is then
\begin{equation}\label{opening-sector}
    -(p+1)\frac{\pi}{2} < \theta < (p+1)\frac{\pi}{2}.
\end{equation}
In view of our earlier conclusions, this was to be expected. Indeed, in analogy to the monomial potential case $V(\phi) = \phi^{2p}$, we expect the asymptotic expansion of $Z(\lambda)$ to be of Gevrey class $1/(p-1)$, and the associated Borel transform $\widehat{\varphi}(\zeta)$ to have a unique pole on the negative real line $\bbR^{-}$. Assuming that $\widehat\varphi(\zeta)$ has the correct exponential size, the Laplace transform $\mathcal{L}_{p-1}\hat\varphi$ is analytic on a sector of opening $(p-1)\pi$ centered on $\bbR^+$; however, it can be analytically continued to a sector of angle $2\pi + (p-1)\pi $ by rotating the integration line of the Laplace transform.
We conclude thus that $Z(\lambda)$ is analytic on the Sokal disc of infinite radius $D^{p-1}_0=\{\lambda \in \bbC: \Re \lambda ^{-\frac{1}{p-1}} > 0\}$. The first assumption of the Nevanlinna-Sokal theorem is then satisfied.

\subsubsection{Gevrey asymptotic expansion}
To apply the Nevanlinna-Sokal theorem we also need to prove that $Z(\lambda)$ admits an asymptotic expansion of Gevrey class $1/(p-1)$. To do so, we use a Taylor expansion with integral remainder: given a function $f(x)$ differentiable infinitely many times at $0$, one has
\begin{equation}
    f(x) = \sum_{k=0}^n\frac{f^{(k)}(0)}{k!}x^k + \frac{x^{n+1}}{n!}\int_0^1du (1-u)^nf^{(n+1)}(xu).
\end{equation}
Hence, after setting $ k! c_k := \frac{\dd^k}{\dd\lambda^k}Z(0)$ we have
\begin{equation}\label{Gevrey-remainder}
    \left|Z(\lambda) - \sum_{k=0}^{n}c_k\lambda^k \right| = \left| \frac{\lambda^{n+1}}{n!}\int_0^1\dd u (1-u)^n\int_{-\infty}^{\infty}\dd\phi e^{-\phi^2-\lambda u V(\phi)}(V(\phi))^{n+1}\right|.
\end{equation}
We now have to prove that the right-hand side can by bounded by $\lambda^{n+1}AB^{n+1}((n+1)!)^{p-1}$ for two constants $A, B$ independent of $\lambda$ and $n$, uniformly on the Sokal disc $D^{p-1}_0=\{\lambda \in \bbC: \Re \lambda ^{-\frac{1}{p-1}} > 0\}$. To do so, we
will perform another change of variable,
$\phi^2 = \lambda^{- (p+1)/p^2}x^2$, and then promote $\lambda$ to a complex variable. The norm can be then reabsorbed in the integration variable be letting $y^2 = |\lambda|^{-(p+1)/p^2}x^2$, whence the second integral in (\ref{Gevrey-remainder}) rewrites as
\begin{align}
    I := \int_{-\infty}^{\infty}\dd y & \,e^{-i \theta\frac{p+1}{2p^2}}\exp\left(-e^{-i\theta\frac{p+1}{p^2}}y^2-u|\lambda|\left[e^{-i\frac{\theta}{p}}y^{2p}+ \ldots \right] \right) V^{n+1}\left(\lambda^{-\frac{p+1}{2p^2}}y\right).
\end{align}
On the Sokal disc $D^{p-1}_0$, we have $|\theta| < (p-1)\frac{\pi}{2}$: thus the integral is well defined because $e^{i \theta/p}$ has positive real part for every $\theta$. In order to bound $|I|$, we can take the integral of the norm of the integrand, which amounts to taking the real part of all the coefficients of the polynomial at the exponential and to replacing $V^{n+1}$ by $|V^{n+1}|$. It will be convenient to separate the terms with positive or negative coefficients at the exponential, by collecting them into two polynomials $f(y)$ and $g(y)$. We can then write
\begin{equation}
    |I| \leq \int_{-\infty}^{\infty}\dd y \,e^{-f(y) + g(y)}h(y),
\end{equation}
where $f(y)$ is a polynomial of degree $2p$ whose coefficients are all positive, and whose leading term is $u |\lambda|\cos\left(\tfrac{\theta}{p}\right)y^{2p}$; $g(y)$ is a polynomial of degree smaller or equal to $2p-1$ whose coefficients are again all positive; finally $h(y) = \left|V^{n+1}\left(y e^{-i\theta (p+1)/(2p^2)}\right)\right|$ is everywhere positive. We will now bound this integral by a Gaussian integral; it will be then easy to find the expected factorial growth of $|I|$. To do so, we split the even and odd powers of $f$ by letting $f(y) = P(y^2) + yQ(y^2)$ and consider the smooth function 
\begin{equation}
    F(c) =   \int_{-\infty}^{\infty}\dd y \left[e^{-f(y) + g(y)} - e^{-cP(y^2)}\right]h(y).
\end{equation}
We then make the following three observations:
\begin{enumerate}
    \item $F(0) = -\infty$, since
    \begin{equation}
        F(0) = I - \int_{-\infty}^{\infty}\dd y \, h(y),
    \end{equation}
    and the second term is divergent to $+\infty$ as $h(y)$ grows polynomially for large $y$;
    \item $F(\infty) = I$;
    \item $F'(c) > 0$, which can be seen by writing \begin{equation}
        F'(c) =   \int_{-\infty}^{\infty}\dd y \left[e^{-f(y) + g(y)} + P(y^2)e^{-cP(y^2)}\right]h(y)
    \end{equation}
and by recalling that $P(y^2)$ is everywhere positive because it has positive coefficients.

\end{enumerate}
Then, by the mean value theorem, there must exist $K>0$ such that $F(K) = 0$ . As $F'(c)>0$, $K$ is unique and $F(c) < 0$ for every positive $c < K$. We are now enabled to write, for some $c < K$,
\begin{equation}
    |I| \leq \int_{-\infty}^{\infty}\dd y\, e^{-cP(y^2)} \left|V^{n+1}\left(y e^{-i\theta \frac{p+1}{2p^2}}\right)\right|
\end{equation}
where $P(y^2) = u|\lambda|\cos\left(\tfrac{\theta}{p}\right)y^{2p}+\ldots$. As all the terms in $P(y^2)$ are positive for every $y$, we can write
\begin{equation}
    e^{-cP(y^2)} \leq e^{-c\cos\left( \theta\frac{p+1}{p^2}\right)y^2}.
\end{equation}
This step enables us to split the integrals in (\ref{Gevrey-remainder}), as now the dependence on $u$ and $|\lambda|$ is dropped. Moreover, as $|\theta| < (p-1)\frac{\pi}{2}$, we have $\cos\left( \theta\frac{p+1}{p^2}\right) > \cos\left( \frac{\pi}{2}\frac{p^2-1}{p^2}\right) =: \varepsilon  $. As it was our intention, we can then bound $|I|$ with the Gaussian integral
\begin{equation}
    |I| \leq  \int_{-\infty}^{\infty} \dd y \,e^{-c\varepsilon y^2} \left|V^{n+1}\left(y e^{-i\theta \frac{p+1}{2p^2}}\right)\right|,
\end{equation}
so that the dependence on $\theta$ of the exponential is also dropped. It is straightforward then to derive 
\begin{equation}\label{final-bound-I}
    |I| \leq (2Ap)^{n+1} \int_{-\infty}^{\infty} \dd y \, e^{-c\varepsilon \;y^2}y^{2p(n+1)} = \frac{1}{\sqrt{c\varepsilon}}\left(\frac{2Ap}{(c\varepsilon)^p}\right)^{n+1}\Gamma\left(p(n+1)+\tfrac{1}{2}\right).
\end{equation}
 where $A:= \max \{|a_i|, i=1,,\ldots,p\}$. The integral over $u$ in (\ref{Gevrey-remainder}) is now trivial and we finally reach a uniform asymptotic expansion is of Gevrey class $1/(p-1)$
\begin{equation}
     \left|Z(\lambda) - \sum_{k=0}^{n}c_k\lambda^k \right| \leq \frac{\lambda^{n+1}}{(n+1)!} \frac{1}{\sqrt{c\varepsilon}}\left(\frac{2Ap}{(c\varepsilon)^p}\right)^{n+1}\Gamma\left(p(n+1)+\tfrac{1}{2}\right)
\end{equation}
for every $\lambda \in D^{p-1}_0$. Thus, the Nevanlinna-Sokal theorem is satisfied and $Z(\lambda)$ is Borel-summable, and more precisely $(p-1)$-summable. Clearly, this proof also captures the monomial potential cases, i.e. $V(\phi) = \phi^{2p}$, dealt with in the previous sections. 
We can conclude that $Z(\lambda)$, when restricted on a finite interval $\lambda \in [0, R] \subset \bbR$,  with $R >0$,  is a tame function, definable in the o-minimal structure $\Rg$.

\subsubsection{Correlation functions}
In analogy to the monomial potential, the correlation functions are also Borel-summable. The function
\begin{equation}
    G_j(\lambda) = \int_{-\infty}^{\infty} \dd \phi \, e^{-\phi^2-\lambda V(\phi)}\phi^j
\end{equation}
with $V(\phi)$ a polynomial as above, can be proved to be Borel-summable following the same steps as before. The only difference is that, instead of (\ref{final-bound-I}), we will get
\begin{equation}
    |I| \leq (2Ap)^{n+1} \int_{-\infty}^{\infty} \dd y\, e^{-c\varepsilon \;y^2}y^{2p(n+1) +j} = \frac{1}{\sqrt{(c\varepsilon)^{j+1}}}\left(\frac{2Ap}{(c\varepsilon)^p}\right)^{n+1}\Gamma\left(p(n+1)+\tfrac{j+1}{2}\right).
\end{equation}
whence uniform Gevrey bounds on the Sokal disc $D^{p-1}_0$ easily follow. Thus, $G_j(\lambda)$ is also $(p-1)$-summable and a tame function in $\Rg$. The correlation functions
\begin{equation}
     \langle\phi^j\rangle = \frac{G_j(\lambda)}{Z(\lambda)}
\end{equation}
are therefore tame functions of $\Rg$, being the ratio of two definable functions of the same structure.

\section{On generalizations to higher dimensions}\label{high-dim}

In the previous section we have used zero-dimensional quantum field theories as a laboratory to gain insight into how the tameness of partition functions and correlation functions persists in the non-analytic weak coupling limits. The power of this setting comes from the availability of non-perturbative results and detailed exact computations. The aim of this section is to discuss extensions of these observations to QFTs formulated in non-zero dimension, and comment on challenges in this endeavour.

\subsection{$\phi^4$ theory in 4 dimensions}
Let us proceed by again considering $\phi^4$ theory, but now in four Euclidean spacetime dimensions, in which the theory is renormalizable. Aside from computational difficulties in evaluating path integrals, the main challenge arises from divergences in the UV and the IR which require regularization. If we were to formulate the theory on a lattice, we would in fact be studying a matrix model in zero dimensions which has already been discussed in the previous section. Another way to implement UV and IR cut-offs, adopted in the constructive field theory literature \cite{Magnen_2008}, is as follows.  We restrict the interaction to a bounded region $U\subseteq \bbR^4$, and write the Schwinger proper time expression for the free propagator,\footnote{In the cited literature, the propagator is thought of as the covariance of the Gaussian distribution of the scalar field.}
\begin{equation}
    \Delta(p) = \int_0^\infty \dd\tau \, e^{-\tau(p^2+m^2)}\,,
\end{equation}
in terms of renormalization group slices indexed by integers $j$,
\begin{equation}
    \Delta_j(p) = \int_{M^{-2j}}^{M^{-2j+2}} \dd\tau \, e^{-\tau(p^2+m^2)}\,.
\end{equation}
Here $M$ parametrizes the width of the slices, and the full propagator is recovered via $\Delta=\sum_j\Delta_j$. A regularized expression for the partition function may then be written as 
\begin{equation}
    Z(\lambda,U) = \int D\phi \, \exp( -\frac{1}{2}\int \dd^4 x\, \phi \Delta^{-1}_j\phi -\lambda \int_U \dd^4 x \,\phi^4)\,.
\end{equation}
Note that only the interaction is localized on the region $U$, since the quadratic term already has an IR cut-off implemented. It was proved in \cite{Magnen_2008}, by means of the Loop Vertex Expansion and the tree formula that the connected correlation functions
\begin{equation}
    S(x_1,\ldots,x_{2p}; \lambda) = \lim_{U \rightarrow \mathbb{R}^4}  \frac{1}{Z(\lambda, U) }\int e^{-\tfrac{1}{2}\int \dd^4 x \phi \Delta^{-1}_j\phi-\lambda\int_U\dd^4x\phi^4}\phi(x_1)\cdots\phi(x_{2p})
\end{equation}
are Borel-summable, uniformly in the slice index $j$. By our general discussion in section \ref{o-minimality-section}, it then follows that $S(x_1,\ldots,x_{2p}; \lambda)$ is a well-defined regularized physical quantity which is definable in the o-minimal structure $\Rg$, despite being non-analytic in the $\lambda\to0$ limit.

\subsection{Vector model in 2 dimensions}
Going beyond theories of a single scalar field, we now consider a theory with $N$ scalars coupled by a general quartic interaction. The Borel summability of such a model was analyzed in \cite{Erbin_2021} for $d=2$, and we will build on these results to argue for the tameness of the partition function. The partition function for the model in question is given by
\begin{equation}
    Z(\lambda) = \int D\phi_1\cdots D\phi_n \, \exp( -\int \dd^2 x \left(\frac{1}{2}\sum_{i=1}^{N} \phi_i \Delta^{-1}\phi_i + \frac{\lambda}{4!} \sum_{i,j,k,l}^{N}\mathcal{W}_{ijkl}\phi_i\phi_j\phi_k\phi_l\right))\,,
\end{equation}
where $\mathcal{W}_{ijkl}$ is a completely symmetric tensor, and $\Delta^{-1}$ is the inverse propagator, which is here regularized by writing it as
\begin{equation}
    \Delta(p) = \int_{M}^\infty \dd\tau \, e^{-\tau(p^2+m^2)}\,.
\end{equation}
where $M$ imposes a UV cut-off. The renormalization of the theory can be implemented by using the so-called Multiscale Loop Vertex Expansion, as detailed in \cite{Gurau:2013oqa}. The authors of \cite{Erbin_2021} proved, under the assumption that $\mathcal{W}_{ijkl}$ has only positive eigenvalues, the largest of which they call $w^2_0$, that the free energy $F(\lambda) = \log Z(\lambda)$ is analytic and Borel-summable in a cardioid domain defined by
\begin{equation}
    |\lambda| \leq O(1)\frac{1}{N w_0^2}\cos^2(\theta/2),
\end{equation}
where $\lambda = |\lambda| e^{i\theta}$ and $O(1)$ denotes constants depending on the specific model. Once more, Borel summability implies that the free energy $F(\lambda)$ is tame in the structure $\Rg$. By composing with the restricted exponential function, we then conclude that the partition function $Z(\lambda)$ is definable in $\Rg$.

\subsection{Tame Gevrey energy eigenvalues in quantum mechanics}\label{qm}
In this final section we wish to discuss an example in quantum mechanics, which may be viewed as a one-dimensional QFT. Here the physical observable of interest will be the energy eigenvalues, in a setting in which the Hamiltonian is perturbed by a term depending on a small coupling $g$. In this case, the energy eigenvalues $E_j(g)$ can in principle be calculated by performing a perturbative asymptotic expansion in $g$.  
In \cite{QM-Borel-summability} the authors considered the time-independent, one-dimensional Hamiltonian
\begin{equation}\label{convex-hamiltonian}
    H(g) = -\frac{1}{2}\frac{\dd^2}{\dd x^2}+ \frac{1}{2}x^2 + \frac{g^{p-1}x^{2p}}{1+\alpha g x^2}\ ,
\end{equation}
where $g$ is the coupling parameter, $\alpha>0$ is a fixed constant and $p\geq 3$ is an integer. Note that this potential is convex: there is only one, global minimum at $x=0$: therefore there are no tunneling phenomena. Theorem 4.1 in \cite{QM-Borel-summability} states that each eigenvalue of the Hamiltonian $E_j(g)$ is analytic in a sector
    \begin{equation}
        S_j = \left\{ g \in \mathbb{C}: 0 < |g| \leq r_j< 1, |\arg g| < (p-1)\frac{\pi}{2}+ \delta\right\}
    \end{equation}
    where $\delta$ is a positive constant. Moreover, every eigenvalue admits an asymptotic expansion
    \begin{equation}
        E_j(g) \sim_{p-1} \sum_{n=0}^{\infty}E_{j,n}g^{n}
    \end{equation}
    namely such that, for every $g \in S_j$, and for every $N$ integer, there exist constants $A_j$ and $B$ such that
\begin{equation}
        \left|E_j(g) - \sum_{n=0}^{N-1}E_{j,n}g^{n}\right|\leq A_j B^N \Gamma\left(N(p-1)+\tfrac{1}{2}\right)|g|^{N}.
\end{equation}

Thus, by the Nevanlinna-Sokal theorem, every eigenvalue $E_j(g)$ is $(p-1)$-summable. The limit for $g \rightarrow 0$ of $E_j(g)$ exists along all directions in $S_j$ and it must be equal to the eigenvalue $j+ \tfrac{1}{2}$ of the harmonic oscillator. Hence, we conclude that every function $E_j(g)$ is a tame function definable in $\Rg$ once restricted on the line $[0,r_j]$.

\section{Conclusions}
In this note we have further supported the idea that physical observables, when considering their dependence on the parameters of the defining Lagrangian, fall in the class of tame functions. This property is motivated by the intuitive demand that they can only capture a finite amount of complexity or logical information. In our analysis we have focused on the tameness principle implemented by o-minimality. More precisely, we have investigated whether the graphs of the physical observables are definable in some o-minimal structure. We have found that this feature can be established for several partition and correlation functions of quantum field theories near their weak coupling limit. This was possible even in the case when observables are non-analytic in this limit, as long as the divergent perturbative series is Borel-summable. The tame  behavior is thus a unifying property beyond analyticity, as already envisioned in \cite{Douglas:2022ynw}. Importantly, we have seen that the relevant o-minimal structures are then $\Rg, \Rge$, the structures defining Gevrey functions \cite{DRIES_SPEISSEGGER_2000}. These structure are strictly larger than the structures $\bbR_{\rm an}$ and $\bbR_{\rm an,exp}$ that have been used in many geometric applications \cite{Bakker2020tame} and in showing that finite-loop amplitudes are tame functions \cite{Douglas:2022ynw}. 

Our explicit examples included 0-dimensional scalar QFTs and later certain classes of vector and matrix models. The latter can be interpreted as lattice QFTs where every site in the lattice is associated with an entry of the matrix. Our results thus provide an important improvement on the findings of \cite{Grimm:2023xqy}, since they also establish the definability in the weak coupling limit of such lattice QFTs. Considering higher-dimensional QFTs on a continuous space-time is a non-trivial task, since the partition and correlation functions are then defined as path integrals which only in rare cases can be evaluated explicitly. Nevertheless, we were able to establish the definability of a regularized partition function for $\phi^4$-theory relying on results from constructive field theory. We also investigated the energy eigenvalues of a quantum mechanical system with a time-independent Hamiltonian perturbed by a convex potential, where we again we demonstrated definability in $\Rg$. Despite these successes, we stress that we have not exploited the structure $\Rg$ in full generality. Throughout this note, we have only been concerned with $p$-summable functions, although the one-variable definable functions of $\Rg$ are only required to be multisummable, a weaker condition. A direction for future investigation is then to understand whether multisummability guarantees the tameness of larger classes of partition functions or energy eigenvalues.

There are at least two important directions in which the program to establish the tameness of QFT observables needs to be extended further. Firstly, it remains open what happens to the tameness of partition and correlation functions that exhibit Stokes phenomena.
In these cases Borel resummation is not sufficient to recover the full observable from the perturbative series \cite{Sauzin:2014qzt}, and generally one then expects that there is additional information that is needed to specify the non-perturbative sector of the theory. Despite this fact, the observables are still strongly constrained by their perturbative expansion and resurgence techniques can be used to infer their properties.
We expect that these conditions can suffice to derive tameness properties, even though one might have to consider an even wider class of o-minimal structures. For example, it appears natural to consider the o-minimal structure constructed in \cite{rolin2007quasi} that combines solutions of differential equations with Borel sums of asymptotic power series. This structure defines functions that have a divergent power series with Stokes phenomena. Generalizations of this type are also needed for quantum mechanical settings with general potentials \cite{vanSpaendonck:2023znn}.
Indeed, if the potential is not convex, the presence of tunneling between potential wells compels us to promote asymptotic perturbative series to transseries, impairing definability in $\Rg$. In this case, however, it is not known if there is a differential equation obeyed by the energy eigenvalues. This further complicates the identification of the o-minimal structure and one might be forced to extend to the general classes constructed in \cite{Rolin_2015}. We hope to return to these questions in future work.

A second key question is to ask if one can find o-minimal structures replacing $\Rg, \Rge$ that admit a notion of complexity. It was suggested in \cite{Grimm:2023xqy} that the required mathematical property is best captured by the notion of sharp o-minimality recently introduced in \cite{beyondomin,binyamini2022sharplyominimalstructuressharp}. However, it is known that $\Rg$ is not sharply o-minimal. This is due to the fact that it contains general restricted analytic functions, which generally contain too much logical information to be compatible with this refined tameness principle.  One way to identify suitable substructures of $\Rg, \Rge$ is to use systems of differential equations obeyed by the partition functions and the correlation functions. If one additionally controls the singularity structure one might hope to generalize the recent remarkable definability results of \cite{newBinyamini} on log-Noetherian functions. It is a far reaching question if there is a unifying sharply o-minimal structure $\bbR_{\rm QFT}$ for a wide class of QFTs. If this is the case the complexity notion can be used as a measure of similarities of QFTs.

Having established that observables in certain QFTs are tame functions and admit a notion of complexity, one can then use the full power of tame geometry to derive rigorous general statements about these settings. We foresee that interesting applications lie in examining symmetries in QFTs and the number of relations these impose on their observables. Complexity can be utilized to identify the simplest fundamental building blocks of observables and to quantify the number of algebraic relations among them. Furthermore, general theorems such as the Pila-Wilkie theorem can provide a compelling connection between the transcendence properties of observables and their symmetries. For instance, one might be able to extend the results of \cite{BKU} (see also \cite{Grimm:2024fip}) beyond the realm of $\mathbb{R}_{\rm an,exp}$, with the expectation that exponential periods will become crucial in this context. Altogether, we are convinced that establishing tameness of observables is the crucial first step to uncovering the universal structures inherent to many QFTs.

\subsubsection*{Acknowledgements}

We would like to thank Mike Douglas, Bruno Klingler, Alexander van Spaendonck, and Marcel Vonk for useful discussions and comments. This research is supported, in part, by the Dutch Research Council (NWO) via a Vici grant.

\appendix

\section{Borel transforms as hypergeometric functions} \label{Borel-explicit}

In this appendix we prove the identities \eqref{hyper-sol} and \eqref{hyper-sol-j}. As for the former, we begin by recalling the most general definition of the hypergeometric function $_aF_b\left[ {\vec{a} \atop \vec{b}} \Big |x\right]$, where $a = |\vec{a}|$, $b = |\vec{b}|$:
\begin{equation}\label{hypergeometric}
  _aF_b\left[ {\vec{a} \atop \vec{b}} \Bigg |x\right] = \dfrac{\prod_{i=1}^{b}\Gamma(b_i)}{\prod_{i=1}^{a}\Gamma(a_i)} \sum_{n=0}^{\infty} \dfrac{\prod_{i=1}^{a}\Gamma(a_i+n)}{\prod_{i=1}^{b}\Gamma(b_i+n)}\frac{x^n}{\Gamma{(n+1)}}.
\end{equation}
We then observe that, for any integer $n$, recalling that $x\Gamma(x) = \Gamma(x+1)$
\begin{align}\label{generalised-recurrence}
  \frac{\Gamma(x+n)}{\Gamma(x)} &=  \frac{\Gamma(x+n)}{\Gamma(x+(n-1))} \frac{\Gamma(x+n-1)}{\Gamma(x+(n-2))} \cdot\cdot\cdot \frac{\Gamma(x+1)}{\Gamma(x)} \\
   \nonumber
   &= (x+n-1)(x+n-2) \cdot\cdot\cdot x .
\end{align}
By comparing (\ref{hyper-sol}) with (\ref{hypergeometric}), we realize that it suffices to prove that
\begin{equation}
  \frac{\Gamma\left( np + \frac{1}{2}\right) }{\Gamma\left(n(p-1)+1\right)} = \sqrt{\pi}\prod_{i=1}^{a}\dfrac{\Gamma(a_i+n)}{\Gamma(a_i)} \left(\prod_{i=1}^{b}\dfrac{\Gamma(b_i+n)}{\Gamma(b_i)}\right)^{-1}\left(\frac{(p-1)^{p-1}}{{p^p}}\right)^n .
\end{equation}
Let us focus on the first product. Plugging in the values in (\ref{abvectors}) and resorting to (\ref{generalised-recurrence}) we have
\begin{alignat}{4} 
\prod_{i=1}^{a}\dfrac{\Gamma(a_i+n)}{\Gamma(a_i)} = \quad
&\big(\tfrac{1}{2p} + n-1\big) && \quad\cdots\quad &&\big(\tfrac{1}{2p}
+1\big) &&\big(\tfrac{1}{2p}\big) \nonumber\\
&\big(\tfrac{3}{2p} + n-1\big) && \quad\cdots\quad && \quad && \big(\tfrac{3}{2p}\big) \nonumber\\
&\quad\vdots && && &&\quad\vdots \nonumber \\
&\big(\tfrac{2p-1}{2p} + n-1\big) && \quad\cdots\quad && \quad && \big(\tfrac{2p-1}{2p}\big).
\end{alignat}

The rows have $n$ terms and the columns have $p$ terms. By taking products starting from the top right corner and moving first down the column and then moving to the column to the left, we have
\begin{equation}\label{earlier-Gamma-hyp}
    \prod_{i=1}^{a}\dfrac{\Gamma(a_i+n)}{\Gamma(a_i)} = \frac{(2np-1)!!}{(2p)^{np}} = \frac{1}{\sqrt{\pi}}\Gamma\left(\frac{2np+1}{2}\right)\frac{1}{p^{np}}
\end{equation}
where we have recalled that $\Gamma(n/2) = \sqrt{\pi}2^{-(n-1)/2}(n-2)!!$ for any odd $n$. Similarly, we can find
\begin{alignat}{4} 
\prod_{i=1}^{b}\dfrac{\Gamma(b_i+n)}{\Gamma(b_i)} = \quad
&\big(\tfrac{1}{p-1} + n-1\big) && \quad\cdots\quad &&\big(\tfrac{1}{p-1}
+1\big) &&\big(\tfrac{1}{p-1}\big) \nonumber\\
&\big(\tfrac{2}{p-1} + n-1\big) && \quad\cdots\quad && \quad && \big(\tfrac{2}{p-1}\big)\nonumber \\
&\quad\vdots && && &&\quad\vdots \nonumber \\
&\big(\tfrac{p-1}{p-1} + n-1\big) && \quad\cdots\quad && \quad && \big(\tfrac{p-1}{p-1}\big) \nonumber
\\[.2cm]
=& \frac{(n(p-1))!}{(p-1)^{n(p-1)}}= \frac{\Gamma(n(p-1)+1)}{(p-1)^{n(p-1)}}
\end{alignat}
which together with the earlier (\ref{earlier-Gamma-hyp}) yields the desired identity.

 The proof of (\ref{hyper-sol-j}) is analogous. The only difference is that, in this case, we have to show that:
\begin{equation}
    \prod_{i=1}^{a}\frac{\Gamma(a_i+n)}{\Gamma(a_i)} = \frac{\Gamma\left(n(p-1)+j+\tfrac{1}{2}\right)}{\Gamma\left(\tfrac{1}{2}+j\right)}.
\end{equation}
The LHS can be expanded as before, yielding the $np$ products
\begin{alignat}{4} 
\prod_{i=1}^{a}\dfrac{\Gamma(a_i+n)}{\Gamma(a_i)} = \quad
&\big(\tfrac{1+2j}{2p} + n-1\big) && \quad...\quad &&\big(\tfrac{1+2j}{2p}
+1\big) &&\big(\tfrac{1+2j}{2p}\big) \nonumber \\
&\big(\tfrac{3+2j}{2p} + n-1\big) && \quad...\quad && \quad && \big(\tfrac{3+2j}{2p}\big) \nonumber\\
&\quad\vdots && && &&\quad\vdots  \nonumber\\
&\big(\tfrac{2p-1+2j}{2p} + n-1\big) && \quad...\quad && \quad && \big(\tfrac{2p-1+2j}{2p}\big).
\end{alignat}
Reading the products from right to left and from the top to the bottom, we realize that the products at the numerator run from $1+2j$ up to $2(np+j)-1$ (the product being only over the odd integers). We conclude that the above product gives the desired result
\begin{equation}
\prod_{i=1}^{a}\dfrac{\Gamma(a_i+n)}{\Gamma(a_i)} = \frac{(2(np+j)-1)!!}{(2j-1)!!}\frac{1}{2^{np}} = \frac{(2(np+j)-1)!!}{2^{np+j}}\frac{2^j}{(2j-1)!!} = \frac{\Gamma\left(np+j+\tfrac{1}{2}\right)}{\Gamma\left(\tfrac{1}{2}+j\right)}\,.
\end{equation}
\section{Gevrey functions of several variables}\label{Rg-many-variables}
In this appendix we provide the full definition of the family $\mathcal{F} = \mathscr{G}$ of Gevrey functions in $m$ variables, which generate the structure $\Rg$. We will need first to fix some notation. If $z \in \bbC$, we conventionally agree that $\arg( z) \in (-\pi ,\pi]$. Let $k =(k_1,..,k_m) \in (0, \infty)^m$, $R= (R_1,\ldots,R_m) \in (0,\infty)^m$ and $z =(z_1,...,z_m) \in \bbC^m$. We will write $z^k = z_1^{k_1}...z_m^{k_m} \in \bbC$, and similarly for $R$. We will write $R < \tilde{R}$ if $ R_i < \tilde{R}_i$ for every $i =1,...,m$ . We will define the product
\begin{equation}
    k \cdot |\arg (z)| = k_1|\arg( z_1)| +\ldots+  k_m|\arg( z_m)|
\end{equation}
and let $|z| = \sup\{|z_i|, i=1,\ldots,m\}$. Let also $[0,R] = [0,R_1]\times \cdots \times [0,R_m]$ and $[0,R)$ similarly.  Finally, we set $\mathscr{K} = \{0\} \cup [1,\infty)$. 

We now define the polydisc $D(R)$ and the polysector $S^k(R, \phi)$, where $\phi \in (0, \pi)$ and $k \in \mathscr{K}^m$, as
\begin{align}
    D(R) &= \{ z \in \bbC^m:| z_i| < R_i \;\textup{for}\; i= 1,\ldots, m\} ,\\ 
    S^k(R, \phi) &= \{ z \in D(R): k\cdot |\arg(z)| < \phi\}.
\end{align}
Notice that, if $m=1$ and $k \geq 1$, one has $S^k(R, \phi) = S(R, \phi, 1/k) \cup \{0\}$ in the notation of section \ref{Rg-section}. For any $p \in \bbN$, let
\begin{align}
    D_p^k(R) &= \left\{ z \in D(R): |z|^k < \frac{R^k}{p+1}\right\}, \\
    S_p^k(R) &= S^k(R, \phi) \cup D_p^k(R).
\end{align}
Finally, for any\textit{ finite}, non-empty set $K \subseteq \mathscr{K}$ we define
\beq
    S^K(R, \phi) = \bigcap_{k \in K}S^k(R, \phi)\, ,\qquad 
    S^K_p(R, \phi) = \bigcap_{k \in K}S^k_p(R, \phi)\, .
\eeq
Given a set $U \subseteq \bbC^m$ and a function $f: U \rightarrow \bbC$, we define the norm
\begin{equation}
   ||f||_U = \sup_{z \in U}\{|f(z)|\} \in [0, \infty].
\end{equation}
It will be useful to define $\tau = (K, R, r, \phi)$, where $K \subseteq \mathscr{K}^m$ is finite and non-empty, $R \in (0, \infty)^m$, $r \in (0, \infty)$, $\phi \in (0, \pi)$. We will then write, for brevity, $S(\tau) = S^K(R, \phi)$ and $S_p(\tau) = S_p^K(R, \phi)$.
Let us fix a certain $\tau$. From our definitions, it is clear that $S_{p+1}(\tau) \subset S_p(\tau)$. Then, if for every $p$, $f_p: S_p(\tau) \rightarrow \bbC$ is a bounded holomorphic function such that $\sum_p ||f_p||_{S_p(\tau)} r^p < \infty$, it can be shown that $\sum_p f_p$ converges to a bounded and continuous function $f: S(\tau) \rightarrow \bbC$. We will denote this fact by 
\begin{equation}
  \sum_p f_p =_\tau f\, .
\end{equation}
We define $\mathscr{G}_\tau$ as the collection of functions $f: S(\tau) \rightarrow \bbC$ such that there exists a sequence $(f_p)_{p \in \bbN}$ of 
functions $S_p(\tau) \rightarrow \bbC$ as above such that $f =_\tau\sum_p f_p$.

Let $\mathscr{T}_m$ to be the collection of all $\tau=(K, R, r, \phi)$, where $K$ and $R$ are as above, while we constrain $r$ and $\phi$ by $r>1$ and $\phi \in (\tfrac{\pi}{2}, \pi)$. A crucial result proved in \cite{DRIES_SPEISSEGGER_2000} is that if $\tau \in \mathscr{T}_m$, $\mathscr{G}_\tau$ is a differential algebra of quasi-analytic functions: i.e.~the Taylor map $\mathscr{G}_\tau \rightarrow \bbC[[Z]]$ 
\begin{equation}
  \mathscr{G}_\tau \ni f \rightarrow \sum_{n \in \bbN^m}\frac{Z^n}{n!}f^{(n)}(0) \in \bbC[[Z]]\, ,
\end{equation}
where $n! = n_1!\cdots n_m!$ and $f^{(n)} = \tfrac{\partial^{n_1}}{\partial z_1^{n_1}}\cdots\tfrac{\partial^{n_m}}{\partial z_m^{n_m}}f$, is injective.

Consider now a polyradius $R = (R_1,\ldots,R_m)$. We define $\mathscr{G}(R)$ to be the ring of functions $f: [0,R] \rightarrow \bbR$ for which there exists $\tilde{\tau} = (\tilde{K}, \tilde{R}, \tilde{r}, \tilde{\phi}) \in \mathscr{T}_m $, with $\tilde{R} > R$, and $\tilde f \in \mathscr{G}_{\tilde \tau}$ such that $f(x) = \tilde{f}(x)$ for every $x \in [0, R]$. The fact that $\mathscr{G}(R)$ is a ring under pointwise addition and multiplication is non-trivial, but we will refer to \cite{DRIES_SPEISSEGGER_2000} for the proof. 

Finally we define $\mathscr{G}$ to be the family of functions $f: \bbR^m \rightarrow \bbR$ which are identically vanishing outside the unit cube $[0,1]^m$ and
\begin{equation}
  f|_{[0,1]^m} \in \mathscr{G}(1,\ldots,1)\,.
\end{equation}
The choice of a polyradius with all entries equal to $1$ is purely conventional and by no means restrictive. We can then define the structure $\Rg$ to be
\begin{equation}
  \Rg = (\bbR,<, 0,1, +, -,\cdot, \mathscr{G})\,.
\end{equation}

\bibliography{references}

\providecommand{\href}[2]{#2}\begingroup\raggedright\begin{thebibliography}{10}

\bibitem{Grimm:2021vpn}
T.~W. Grimm, ``{Taming the landscape of effective theories},''
  \href{http://dx.doi.org/10.1007/JHEP11(2022)003}{{\em JHEP} {\bfseries 11}
  (2022) 003}, \href{http://arxiv.org/abs/2112.08383}{{\ttfamily
  arXiv:2112.08383 [hep-th]}}.

\bibitem{Douglas:2022ynw}
M.~R. Douglas, T.~W. Grimm, and L.~Schlechter, ``{The Tameness of Quantum Field
  Theory, Part I -- Amplitudes},''
  \href{http://arxiv.org/abs/2210.10057}{{\ttfamily arXiv:2210.10057
  [hep-th]}}.

\bibitem{Douglas:2023fcg}
M.~R. Douglas, T.~W. Grimm, and L.~Schlechter, ``{The Tameness of Quantum Field
  Theory, Part II -- Structures and CFTs},''
  \href{http://arxiv.org/abs/2302.04275}{{\ttfamily arXiv:2302.04275
  [hep-th]}}.

\bibitem{Grimm:2023xqy}
T.~W. Grimm, L.~Schlechter, and M.~van Vliet, ``{Complexity in tame quantum
  theories},'' \href{http://dx.doi.org/10.1007/JHEP05(2024)001}{{\em JHEP}
  {\bfseries 05} (2024) 001}, \href{http://arxiv.org/abs/2310.01484}{{\ttfamily
  arXiv:2310.01484 [hep-th]}}.

\bibitem{MR1633348}
L.~van~den Dries, {\em Tame topology and o-minimal structures}, vol.~248 of
  {\em London Mathematical Society Lecture Note Series}.
\newblock Cambridge University Press, Cambridge, 1998.

\bibitem{beyondomin}
G.~Binyamini and D.~Novikov, {\em Tameness in geometry and arithmetic: beyond
  o-minimality}, \href{http://dx.doi.org/10.4171/icm2022/117}{pp.~1440--1461}.
\newblock 12, 2023.

\bibitem{binyamini2022sharplyominimalstructuressharp}
G.~Binyamini, D.~Novikov, and B.~Zack, ``Sharply o-minimal structures and sharp
  cellular decomposition,'' \href{http://arxiv.org/abs/2209.10972}{{\ttfamily
  arXiv:2209.10972 [math.LO]}}.

\bibitem{newBinyamini}
G.~Binyamini, ``{Log-Noetherian functions},''
  \href{http://arxiv.org/abs/2405.16963}{{\ttfamily arXiv:2405.16963
  [math.AG]}}.

\bibitem{PhysRev.85.631}
F.~J. Dyson, ``Divergence of perturbation theory in quantum electrodynamics,''
  {\em Phys. Rev.} {\bfseries 85} (Feb, 1952) 631--632.

\bibitem{DRIES_SPEISSEGGER_2000}
L.~v.~d. Dries and P.~Speisegger, ``The field of reals with multisummable
  series and the exponential function,''
  \href{http://dx.doi.org/10.1112/S0024611500012648}{{\em Proceedings of the
  London Mathematical Society} {\bfseries 81} no.~3, (2000) 513–565}.

\bibitem{DenjoyCarleman}
J.~Rolin, P.~Speissegger, and A.~Wilkie, ``{Quasianalytic Denjoy–Carleman
  Classes and O-Minimality},''
  \href{http://dx.doi.org/10.1090/S0894-0347-03-00427-2}{{\em Journal of the
  American Mathematical Society} {\bfseries 16} (10, 2003) }.

\bibitem{Sauzin:2014qzt}
D.~Sauzin, ``{Introduction to 1-summability and resurgence},''
  \href{http://arxiv.org/abs/1405.0356}{{\ttfamily arXiv:1405.0356 [math.DS]}}.

\bibitem{balser1994divergent}
W.~Balser, {\em From Divergent Power Series to Analytic Functions}, vol.~1582
  of {\em Lecture Notes in Mathematics}.
\newblock Springer, 1994.

\bibitem{Hardy-book}
G.~H. Hardy, {\em {Divergent Series}}.
\newblock Clarendon Press, Oxford, 1956.

\bibitem{LodayRichaud2014DivergentSA}
M.~Loday-Richaud, ``Divergent series and differential equations,''
\newblock 2014.
\newblock \url{https://api.semanticscholar.org/CorpusID:227021907}.

\bibitem{Sokal-improvemnt-on-Watson}
A.~Sokal, ``An improvement of {W}atson's theorem on borel summability,''
  \href{http://dx.doi.org/10.1063/1.524408}{{\em Journal of Mathematical
  Physics} {\bfseries 21} no.~2, (1979) 261--263}.

\bibitem{Magnen_2009}
J.~Magnen, K.~Noui, V.~Rivasseau, and M.~Smerlak, ``{Scaling behaviour of
  three-dimensional group field theory},'' {\em Class. Quant. Grav.} {\bfseries
  26} (2009) 185012, \href{http://arxiv.org/abs/0906.5477}{{\ttfamily
  arXiv:0906.5477 [hep-th]}}.

\bibitem{Rivasseau:2023qzm}
V.~Rivasseau, ``{Loop Vertex Representation for Cumulants},''
  \href{http://arxiv.org/abs/2305.08399}{{\ttfamily arXiv:2305.08399
  [math-ph]}}.

\bibitem{Ferdinand:2022duk}
L.~Ferdinand, R.~Gurau, C.~I. Perez-Sanchez, and F.~Vignes-Tourneret, ``{Borel
  Summability of the ${\textrm{1}/N}$ Expansion in Quartic
  ${\textrm{O}(N)}$-Vector Models},''
  \href{http://dx.doi.org/10.1007/s00023-023-01350-w}{{\em Annales Henri
  Poincare} {\bfseries 25} no.~3, (2024) 2037--2064},
  \href{http://arxiv.org/abs/2209.09045}{{\ttfamily arXiv:2209.09045
  [math-ph]}}.

\bibitem{Lionni:2016ush}
L.~Lionni and V.~Rivasseau, ``{Intermediate Field Representation for Positive
  Matrix and Tensor Interactions},''
  \href{http://dx.doi.org/10.1007/s00023-019-00833-z}{{\em Annales Henri
  Poincare} {\bfseries 20} no.~10, (2019) 3265--3311},
  \href{http://arxiv.org/abs/1609.05018}{{\ttfamily arXiv:1609.05018
  [math-ph]}}.

\bibitem{Grimm:2024mbw}
T.~W. Grimm, A.~Hoefnagels, and M.~van Vliet, ``{Structure and Complexity of
  Cosmological Correlators},''
  \href{http://arxiv.org/abs/2404.03716}{{\ttfamily arXiv:2404.03716
  [hep-th]}}.

\bibitem{Rivasseau_2009}
V.~Rivasseau, ``Constructive field theory in zero dimension,'' {\em Advances in
  Mathematical Physics} (6, 2009) 1–12,
  \href{http://arxiv.org/abs/0906.3524}{{\ttfamily arXiv:0906.3524 [math-ph]}}.

\bibitem{Fauvet_2020}
F.~Fauvet, F.~Menous, and J.~Qu\'eva, ``{Resurgence and holonomy of the
  $\phi^{2k}$ model in zero dimension},''
  \href{http://dx.doi.org/10.1063/5.0009292}{{\em J. Math. Phys.} {\bfseries
  61} no.~9, (2020) 092301}, \href{http://arxiv.org/abs/1910.01606}{{\ttfamily
  arXiv:1910.01606 [math-ph]}}.

\bibitem{Rivasseau_2017}
V.~Rivasseau, ``Loop vertex expansion for higher-order interactions,'' {\em
  Letters in Mathematical Physics} {\bfseries 108} no.~5, (Dec., 2017)
  1147–1162, \href{http://arxiv.org/abs/1702.07602}{{\ttfamily
  arXiv:1702.07602 [math-ph]}}.

\bibitem{Rivasseau_2007}
V.~Rivasseau, ``{Constructive Matrix Theory},''
  \href{http://dx.doi.org/10.1088/1126-6708/2007/09/008}{{\em JHEP} {\bfseries
  09} (2007) 008}, \href{http://arxiv.org/abs/0706.1224}{{\ttfamily
  arXiv:0706.1224 [hep-th]}}.

\bibitem{abdesselam1995trees}
A.~Abdesselam and V.~Rivasseau, ``Trees, forests and jungles: A botanical
  garden for cluster expansions,'' {\em Lecture Notes in Physics} {\bfseries
  446} (1995) 7--36.

\bibitem{Rivasseau_2013}
V.~Rivasseau and Z.~Wang, ``{How to Resum Feynman Graphs},''
  \href{http://dx.doi.org/10.1007/s00023-013-0299-8}{{\em Annales Henri
  Poincare} {\bfseries 15} no.~11, (2014) 2069--2083},
  \href{http://arxiv.org/abs/1304.5913}{{\ttfamily arXiv:1304.5913 [math-ph]}}.

\bibitem{Krajewski:2017thd}
T.~Krajewski, V.~Rivasseau, and V.~Sazonov, ``{Constructive Matrix Theory for
  Higher Order Interaction},''
  \href{http://dx.doi.org/10.1007/s00023-019-00845-9}{{\em Annales Henri
  Poincare} {\bfseries 20} no.~12, (2019) 3997--4032},
  \href{http://arxiv.org/abs/1712.05670}{{\ttfamily arXiv:1712.05670
  [math-ph]}}.

\bibitem{Krajewski:2019tsi}
T.~Krajewski, V.~Rivasseau, and V.~Sazonov, ``{Constructive Matrix Theory for
  Higher Order Interaction II: Hermitian and Real Symmetric Cases},''
  \href{http://dx.doi.org/10.1007/s00023-022-01170-4}{{\em Annales Henri
  Poincare} {\bfseries 23} no.~10, (2022) 3431--3452},
  \href{http://arxiv.org/abs/1910.13261}{{\ttfamily arXiv:1910.13261
  [math-ph]}}.

\bibitem{Magnen_2008}
J.~Magnen and V.~Rivasseau, ``{Constructive $\phi^4$ Field Theory without
  Tears},'' {\em Annales Henri Poincaré} {\bfseries 9} no.~2, (Apr., 2008)
  403–424, \href{http://arxiv.org/abs/0706.2457}{{\ttfamily arXiv:0706.2457
  [math-ph]}}.

\bibitem{Erbin_2021}
H.~Erbin, V.~Lahoche, and M.~Tamaazousti, ``{Constructive expansion for vector
  field theories I. Quartic models in low dimensions},''
  \href{http://dx.doi.org/10.1063/5.0038599}{{\em J. Math. Phys.} {\bfseries
  62} no.~4, (2021) 043501}, \href{http://arxiv.org/abs/1904.05933}{{\ttfamily
  arXiv:1904.05933 [hep-th]}}.

\bibitem{Gurau:2013oqa}
R.~Gurau and V.~Rivasseau, ``{The Multiscale Loop Vertex Expansion},''
  \href{http://dx.doi.org/10.1007/s00023-014-0370-0}{{\em Annales Henri
  Poincare} {\bfseries 16} no.~8, (2015) 1869--1897},
  \href{http://arxiv.org/abs/1312.7226}{{\ttfamily arXiv:1312.7226 [math-ph]}}.

\bibitem{QM-Borel-summability}
M.~Gomes, ``Borel-leroy summability of a nonpolynomial potential,''
  \href{http://dx.doi.org/10.1016/S0034-4877(08)80021-3}{{\em Reports on
  Mathematical Physics - REP MATH PHYS} {\bfseries 61} (06, 2008) 401--415}.

\bibitem{Bakker2020tame}
B.~Bakker, B.~Klingler, and J.~Tsimerman, ``Tame topology of arithmetic
  quotients and algebraicity of hodge loci,'' {\em Journal of the American
  Mathematical Society} {\bfseries 33} no.~4, (2020) 917--939,
  \href{http://arxiv.org/abs/1803.09384}{{\ttfamily arXiv:1803.09384
  [math.AG]}}.

\bibitem{rolin2007quasi}
J.-P. Rolin, F.~Sanz, and R.~Sch{\"a}fke, ``Quasi-analytic solutions of
  analytic ordinary differential equations and o-minimal structures,'' {\em
  Proceedings of the London Mathematical Society} {\bfseries 95} no.~2, (2007)
  413--442, \href{http://arxiv.org/abs/math/0505073}{{\ttfamily
  arXiv:math/0505073}}.

\bibitem{vanSpaendonck:2023znn}
A.~van Spaendonck and M.~Vonk, ``{Exact instanton transseries for quantum
  mechanics},'' \href{http://arxiv.org/abs/2309.05700}{{\ttfamily
  arXiv:2309.05700 [hep-th]}}.

\bibitem{Rolin_2015}
J.-P. Rolin and T.~Servi, ``Quantifier elimination and rectilinearization
  theorem for generalized quasianalytic algebras,'' {\em Proceedings of the
  London Mathematical Society} {\bfseries 110} no.~5, (Mar., 2015) 1207–1247,
  \href{http://arxiv.org/abs/1303.3724}{{\ttfamily arXiv:1303.3724 [math.AG]}}.

\bibitem{BKU}
G.~Baldi, B.~Klingler, and E.~Ullmo, ``{On the distribution of the Hodge
  locus},'' {\em Inventiones mathematicae} {\bfseries 235} no.~2, (2024)
  441--487, \href{http://arxiv.org/abs/2107.08838}{{\ttfamily arXiv:2107.08838
  [math.AG]}}.

\bibitem{Grimm:2024fip}
T.~W. Grimm and D.~van~de Heisteeg, ``{Exact Flux Vacua, Symmetries, and the
  Structure of the Landscape},''
  \href{http://arxiv.org/abs/2404.12422}{{\ttfamily arXiv:2404.12422
  [hep-th]}}.

\end{thebibliography}\endgroup
\bibliographystyle{utphys}

\end{document}